\documentclass[aps,prd,twocolumn,amsmath,superscriptaddress,amssymb,showpacs,floatfix,nofootinbib,longbibliography]{revtex4-1}

\usepackage{graphicx}
\usepackage{dcolumn}
\usepackage{xcolor}
\usepackage{bm}
\usepackage{natbib}
\usepackage{calligra}
\usepackage[T1]{fontenc}
\usepackage{egothic}
\usepackage[T1]{fontenc}
\newfont{\rsfsten}{rsfs10 scaled 1200}
\newfont{\rsfsseven}{rsfs10 scaled 1200}
\newfont{\rsfsfive}{rsfs10 scaled 1200}
\usepackage{epsfig}
\usepackage{units}
\usepackage[utf8]{inputenc}
\usepackage{hyperref}

\newcommand{\aap}{{Astron.~Astrophys.}}

\newcommand{\apjs}{{Astrophys.~J.~Supp.}}

\newcommand{\mnras}{{Mon.~Not.~R.~Astron.~Soc.}}

\newcommand{\apss}{{Astrophys.~and~Space~Science}}

\newcommand{\be}{\begin{equation}}
\newcommand{\ee}{\end{equation}}
\newcommand{\bea}{\begin{eqnarray}}
\newcommand{\eea}{\end{eqnarray}}

\def\lsim{\mathrel{\raise.3ex\hbox{$<$\kern-.75em\lower1ex\hbox{$\sim$}}}}
\def\gsim{\mathrel{\raise.3ex\hbox{$>$\kern-.75em\lower1ex\hbox{$\sim$}}}}

\begin{document}

\title{Scrutinizing the Isotropic Gamma-Ray Background in Search of Dark Matter}

\author{Ilias Cholis}
\email{cholis@oakland.edu, ORCID: orcid.org/0000-0002-3805-6478}
\affiliation{Department of Physics, Oakland University, Rochester, Michigan, 48309, USA}
\author{Iason Krommydas}
\email{ik23@rice.edu, ORCID: orcid.org/0000-0001-7849-8863}
\affiliation{Department of Physics and Astronomy, Rice University, Houston, Texas, 77005, USA}
\date{\today}

\begin{abstract}
The isotropic gamma-ray background (IGRB), measured by the \textit{Fermi} Large Area Telescope, is the result of several classes of extragalactic astrophysical sources. 
Those sources include blazars, start-forming galaxies and radio galaxies. 
Also, ultra-high-energy cosmic rays interacting with the infrared background, contribute to the isotropic background. 
Using information from \textit{Fermi}'s gamma-ray sources catalog and the results of dedicated studies of these classes of sources, from  observations at the infrared and radio, we model their contribution to the IGRB. In addition to conventional astrophysical sources, dark matter may be a component of the IGRB. 
We combine our model of conventional astrophysical sources and of dark matter annihilation in distant galaxies, marginalizing over relevant uncertainties, to derive constraints on the dark matter annihilation cross section, from the measured IGRB.
In calculating the contribution from dark matter we include the flux from extragalactic halos and their substructure and also the subdominant contribution from Milky Way's halo at high galactic latitudes.
The resulting constraints are competitive with the strongest current constraints from the dwarf spheroidal galaxies. Under certain dark matter assumptions, we also find an indication for a small excess flux in the isotropic background. Our results are consistent with the gamma-ray excess at GeV energies toward the galactic center.

\end{abstract}

\maketitle

\section{Introduction}
\label{sec:introduction}

The isotropic gamma-ray background (IGRB), is the combined emission of a vast number of unresolved sources. 
These sources include active galactic nuclei (AGN)~\cite{Stecker:1993ni,1995MNRAS.277.1477P,  Salamon:1994ku,Stecker:1996ma,Mukherjee:1999it,Narumoto:2006qg,Giommi:2005bp,Dermer:2006pd,Pavlidou:2007dv, Inoue:2008pk, Ajello:2013lka, 
DiMauro:2013zfa, Qu:2019zln, Zeng:2021wln, Korsmeier:2022cwp},
radio galaxies~\cite{2011ApJ...733...66I, 
Stecker:2010di, Inoue:2011bm, DiMauro:2013xta, Hooper:2016gjy, Fukazawa:2022gwm} and star forming galaxies~\cite{Pavlidou:2002va,Thompson:2006qd,Fields:2010bw,Makiya:2010zt, Stecker:2010di, Chakraborty:2012sh, Lacki:2012si, Tamborra:2014xia, Fornasa:2015qua, Linden:2016fdd, Blanco:2021icw}. 
Also, unresolved galactic gamma-ray sources as recycled pulsars located on the sky far away from the galactic disk, can contribute to the IGRB \cite{2011MNRAS.415.1074S, Gregoire:2013yta, Cholis:2013ena}. 
In addition to those sources ultra-high-energy cosmic rays (UHECR) propagating through the intergalactic medium and 
interacting with low-energy photons give cascades of particles that include gamma rays \cite{1973Natur.241..109S, 1973Ap&SS..20...47S, Cholis:2013ena, Globus:2017ehu, Aloisio:2017iyh, AlvesBatista:2019rhs}. 
Finally, dark matter in distant galaxies and in the Milky Way, under certain assumptions on its mass-scale and coupling to 
standard model particles can contribute to the measured IGRB (see e.g.~\cite{Siegal-Gaskins:2013tga, Ando:2013ff, Cholis:2013ena, Bringmann:2013ruh, Ajello:2015mfa, DiMauro:2015tfa}). 

As longer observations of the gamma-ray sky are performed, more of the previously unresolved point sources are identified.
This reduces the flux of gamma rays defined as the IGRB, from its first observation by the \textit{SAS-2} satellite~\cite{1978ApJ...222..833F}, 
then by \textit{EGRET}~\cite{EGRET:1997qcq}, and currently by the \textit{Fermi} Large Area Telescope (LAT)~\cite{Fermi-LAT:2010pat}.
With its much higher sensitivity \textit{Fermi}-LAT has allowed us to perform detailed point-sources analyses as \cite{Malyshev:2011zi, Fermi-LAT:2015otn, Zechlin:2015wdz, Lisanti:2016jub}, that then provided us with an unprecedented level of modeling on the extragalactic and galactic 
gamma-ray sources. Refs.~\cite{Abazajian:2010pc, 2011ApJ...743..171A, Fermi-LAT:2015otn, Manconi:2019ynl}, relying on the identification of 100s of AGNs, 
have derived the luminosity 
functions and redshift distributions of flat-spectrum radio quasars (FSRQs) and BL Lacertae (BL Lac) objects. The combined contribution from
these sources is expected to be 20-30$\%$ of the IGRB spectrum~\cite{Fermi-LAT:2010tsy, Ajello:2011zi, Cuoco:2012yf, Harding:2012gk, Cholis:2013ena, Fornasa:2015qua, DiMauro:2015tfa}. 
In turn, star-forming galaxies, star-burst galaxies and radio galaxies are expected to explain most
of the IGRB emission~\cite{Stecker:2010di, Fermi-LAT:2012nqz, Inoue:2011bm, Cavadini:2011ig, Cholis:2013ena, Linden:2016fdd, Hooper:2016gjy, Blanco:2021icw}. 
To model the luminosity and redshift distributions of 
these galaxies, correlations between gamma-ray and infrared or radio emission have been performed. UHECRs possibly contribute 
significantly at the higher-end of the IGRB spectrum~\cite{Ahlers:2011sd, Gelmini:2011kg, Cholis:2013ena, Globus:2017ehu, Aloisio:2017iyh, AlvesBatista:2019rhs}.
However, each population of gamma-ray sources has some remaining modeling uncertainties. In this work, we explore the 
potential contribution from annihilating dark matter to the IGRB spectrum. 

In section~\ref{sec:data}, we describe the \textit{Fermi}-LAT IGRB spectral data used in this analysis. Then in section~\ref{sec:astro_backgrounds},
we present how each population of known astrophysical components to the IGRB is modeled; and also how we then combine these 
separate components to interpret the observed IGRB spectrum with its uncertainties, In section~\ref{sec:DM}, we describe how we 
model the contribution from dark matter annihilations to the IGRB. We calculate separately the dominant component from the cumulative 
emission from distant galaxies including their substructures and the smaller component emission from annihilations inside the Milky Way.
In section~\ref{sec:DMlimits}, we present the upper limits on dark matter annihilation cross-section with mass between 5 GeV and 5 TeV 
for a collection of simple annihilation channels to standard model particles. 
We find that the current IGRB spectrum can not exclude a thermal relic dark matter particle and that under certain modeling assumptions there is a slight statistical preference for a dark matter component. 
Our results depend very little on the exact extragalactic dark matter halo mass distribution (the halo mass function). Instead, they are more susceptible at a factor of $\simeq 2$, to the systematic uncertainties of the evaluated IGRB spectrum. 

We also discuss how our limits relate to the 
observations of the Galactic Center and the Inner Galaxy where an excess of gamma-ray emission, known as the Galactic Center Excess (GCE) 
has been firmly detected~\cite{Goodenough:2009gk, Hooper:2010mq, Abazajian:2010zy, Hooper:2011ti, Hooper:2013rwa, Gordon:2013vta, Abazajian:2014fta, Daylan:2014rsa, Calore:2014xka, Zhou:2014lva, Fermi-LAT:2015sau, Linden:2016rcf, Macias:2019omb, DiMauro:2021raz, Cholis:2021rpp}. 
That excess can possibly be explained as a signal of annihilating dark matter \cite{Calore:2014nla, Agrawal:2014oha, Berlin:2015wwa, Karwin:2016tsw, Fermi-LAT:2017opo, Leane:2019xiy}, even with our most current understanding of the GCE spectral and
morphological properties \cite{Cholis:2021rpp, DiMauro:2021raz, Zhong:2024vyi}. Alternative astrophysical explanations for the GCE
have been proposed in literature, with an active debate on their compatibility to gamma-ray observations unrelated to the inner galaxy and at other wavelengths
\cite{Abazajian:2012pn, Hooper:2013nhl, Abazajian:2014fta, Petrovic:2014xra, Cholis:2014lta, Petrovic:2014uda, Carlson:2014cwa, Cholis:2015dea, Bartels:2015aea, Lee:2015fea, Brandt:2015ula, Hooper:2016rap, Bartels:2017vsx, Haggard:2017lyq, Macias:2019omb, Buschmann:2020adf, Gautam:2021wqn}, and with varying levels of compatibility to the more recent analyses on properties of the GCE~\cite{Zhong:2019ycb, Leane:2020nmi, Leane:2020pfc, Cholis:2021rpp, McDermott:2022zmq, Zhong:2024vyi} (see however \cite{Macias:2019omb, Pohl:2022nnd, Song:2024iup}).
We also compare our results to limits from searches toward 
or dwarf spheroidal galaxies~\cite{HESS:2010epq, Cholis:2012am, Fermi-LAT:2013sme, Geringer-Sameth:2014qqa, Geringer-Sameth:2015lua, Fermi-LAT:2015att, Hooper:2015ula, Li:2015kag, Fermi-LAT:2016uux, MAGIC:2016xys, Strigari:2018utn, Boddy:2018qur, Calore:2018sdx, HESS:2020zwn, Alvarez:2020cmw}. 
Finally in section~\ref{sec:conclusions}, we give our conclusions and discuss future improvements that can allow us to reduce 
astrophysical systematic uncertainties associated to the modeling of the diffuse emission from our galaxy. Such improvements 
can provide tighter limits on annihilating dark matter or more robustly identify any spectral features on the IGRB.   

\section{The Isotropic Gamma-Ray Background Spectrum}
\label{sec:data}

We use the \textit{Fermi}-LAT IGRB spectrum of Ref.~\cite{Fermi-LAT:2014ryh}.
That spectrum excludes the emission from high galactic latitude resolved point sources and also removes the 
galactic diffuse emission contribution at high latitudes. The latter contribution is 
model dependent which results in foreground model uncertainties. Ref.~\cite{Fermi-LAT:2014ryh}, used three alternative 
models for the foreground galactic diffuse emission to account for uncertainties in the galactic cosmic-ray distribution and
spectrum, and the properties of the interstellar medium related the interstellar gas and the interstellar radiation field at high latitudes. 

In this analysis, we use the IGRB spectrum from 100 MeV to 820 GeV evaluated when using all three foreground 
galactic emission models; models ``A'', ``B'' and ``C'' from Ref.~\cite{Fermi-LAT:2014ryh}. Model ``B'' predicts the highest IGRB flux 
in the entire energy range, while model ``A'' gives the lowest flux in the in the more important for this analysis $\simeq 1-100$ 
GeV range where the IGRB errors are the smallest. Model ``C'' provides at first glance 
a similar IGRB spectrum as model ``A''.  
We used for each of these IGRB fluxes the reported IGRB spectral errors. By testing our results on all three evaluations of the 
IGRB, we account for the foreground related uncertainties. We note that in subtracting the galactic diffuse emission, to get the IGRB 
spectra, the possible contribution from dark matter annihilations taking place in the Milky Way at high latitudes is not accounted for.
Thus, such a component has to be included when evaluating the dark matter contribution to the IGRB, as we describe in Sec.~\ref{sec:DM}. 

\section{The contribution to the IGRB from known astrophysical sources}
\label{sec:astro_backgrounds}

The IGRB is the result of the combined emission from extragalactic and galactic gamma-ray sources 
that are not bright enough to be identified as either point or extended sources by the \textit{Fermi}-LAT 
collaboration in its relevant catalogues \cite{Fermi-LAT:2009ihh, Fermi-LAT:2011iqa, Fermi-LAT:2015bhf, 
Fermi-LAT:2019yla, Ballet:2023qzs}. As the IGRB is evaluated at galactic latitudes of $|b|> 20^{\circ}$,
after subtracting the galactic diffuse foreground emission \cite{Fermi-LAT:2014ryh}, most of its intensity 
is of extragalactic origin. However, some galactic gamma-ray point sources, below the detection threshold 
may contribute. 

Following past work as \cite{Bringmann:2013ruh, Cholis:2013ena, DiMauro:2015tfa, Ajello:2015mfa, Fornasa:2015qua}, 
we break the known astrophysical gamma-ray sources contributing to the IGRB into 
five components. The contribution from blazars, subdivided BL Lac objects and FSRQs; from star-forming galaxies, from radio galaxies, gamma-rays 
produced from  the interaction of UHECRs with the intergalactic medium and 
gamma rays emitted from Millisecond Pulsars (MSPs). The first four components are extragalactic, while 
MSPs are galactic sources that due to high natal kicks have escaped the galactic disk or the stellar clusters they may have 
originated from but are still gravitationally bound to the Milky Way. 
We model each component separately and then fit the combination of all five components to the IGRB spectrum.  

\subsection{Modeling the contribution from BL Lacertae objects and Flat-Spectrum Radio Quasars}
\label{subsec:BLLac-FSRQ}

Active galactic nuclei (AGN) represent the most numerous class of resolved extragalactic gamma-ray sources 
(see e.g.~\cite{Fermi-LAT:2019yla}), with BL Lacs and FSRQs being the two basic types of AGNs. Such resolved 
objects do not contribute to the IGRB. Thus, we focus here only on the objects that lie below the \textit{Fermi} 
detection threshold. We rely on the fact that both these classes have large number of detected objects, with 
measured gamma-ray spectra, at different redshifts and have also been detected at other wavelengths. This 
allows us to build well-grounded redshift and luminosity distributions for these objects (see e.g.~\cite{Ajello:2013lka, 
DiMauro:2013zfa, Qu:2019zln, Zeng:2021wln, Korsmeier:2022cwp}). 

Following the recent analysis of~\cite{Korsmeier:2022cwp}, we assume that the BL Lacs and FSRQs 
have a gamma-ray luminosity function that can be parametrized as, 
\begin{equation}
\Phi (L_{\gamma}, \Gamma, z) = \Phi (L_{\gamma}, \Gamma, 0) \times e(L_{\gamma}, z). 
\label{eq:AGN_GLF}
\end{equation}
The gamma-ray luminosity function gives the number of sources per range of luminosity $L_{\gamma}$, redshift $z$, 
co-moving volume $V$ and photon spectral index $\Gamma$;
$\Phi (L_{\gamma}, \Gamma, z) = d^{3}N/dL_{\gamma} dV d\Gamma$.
The luminosity $L_{\gamma}$ is defined in the source's rest frame within the range of 0.1-100 GeV and the 
differential spectrum of the source in gamma rays is $dN/dE_{\gamma} \propto E_{\gamma}^{\Gamma}$ 
(see Ref.~\cite{Korsmeier:2022cwp} for further details). 

At redshift of $z=0$, the gamma-ray luminosity function is,
\begin{eqnarray}
\Phi (L_{\gamma}, \Gamma, 0) &=& \frac{A}{ln(10)L_{\gamma}}
 \left[ \left( \frac{L_{\gamma}}{L_{0}} \right)^{\gamma_{1}} + 
 \left( \frac{L_{\gamma}}{L_{0}} \right)^{\gamma_{2}} \right]^{-1} \nonumber \\
 &\times& exp \left[ - \frac{(\Gamma - \mu(L_{\gamma}))^{2}}{2 \sigma^2}\right]. 
\label{eq:AGN_GLF_z0}
\end{eqnarray}
$A$ is a relevant normalization factor, $\gamma_{1, 2}$ indices give the relevant
dependence of the $\Phi (L_{\gamma}, \Gamma, 0)$ on $L_{\gamma}$ around a pivot value $L_{0}$. 
Sources are
taken to follow a Gaussian distribution of spectral indices, with a dispersion $\sigma$
around a mean $\mu$, that depends on the luminosity of the source,
\begin{equation}
\mu (L_{\gamma}) = \mu^{*}  + \beta \left[ log_{10}\left( \frac{L_{\gamma}}{\textrm{erg s}^{-1}}\right) - 46 \right]. 
\label{eq:AGN_mu_Lgamma}
\end{equation}
BL Lacs and SFQRs have the same parameterization of Eq.~\ref{eq:AGN_GLF}-\ref{eq:AGN_mu_Lgamma}, 
but with their own values for each parameter fitted to the observed \textit{Fermi} catalogue. We use the 
values of Ref.~\cite{Korsmeier:2022cwp} that have been derived from their 4FGL catalogue fit (see their 
Table 2); with the exception that when doing our fits of the contribution from all types of sources to the IGRB 
spectrum, we find for the BL Lacs a better fit for $\mu^{*} = 2.3$ instead of 2.03. For the FSRQs we still used 
the $\mu^{*} = 2.5$ suggested by \cite{Korsmeier:2022cwp}. However, as we describe in Sec.~\ref{subsec:analysis}, 
in fitting to the IGRB spectrum, we allow for the spectral index of both the BL Lac and the FSRQ components 
to have some level of freedom. This translates to a freedom in the value of $\mu^{*}$ of these objects. 

All extragalactic gamma-ray sources have attenuation to their observed spectra due to interactions of 
gamma rays with the intergalactic background light leading to electron-positron pair production. We use 
two alternative models for the
optical depth $\tau(E_{\gamma}, z)$ provided in Ref.~\cite{2012MNRAS.422.3189G} as ``fiducial'' and ``fixed''. 
We use those two models for all extragalactic sources of gamma rays, including dark matter and present relevant 
results.

The total intensity of gamma rays from a class of sources is in turn,
\begin{eqnarray}
I(E_{\gamma}) &=& \int_{0}^{z_{\textrm{max}}}dz \int_{\Gamma_{\textrm{min}}}^{\Gamma_{\textrm{max}}} d\Gamma
\int_{L_{\gamma}^{\textrm{min}}}^{L_{\gamma}^{\textrm{max}}} \frac{dV}{dz} \nonumber \\
&\times& \Phi (L_{\gamma}, \Gamma, z) \cdot \frac{dN}{dE}(E_{\gamma}(1+z)) \cdot \frac{L_{\gamma}}{2 \pi (d_{L}(z))^{2}} \nonumber \\
&\times& exp \left[ - \tau(z, E_{\gamma}(1+z)) \right] \cdot (1-\omega_{\gamma}(L_{\gamma},z)),
\label{eq:I_AGN}
\end{eqnarray}
where $E_{\gamma}$ refers to the gamma-ray energy at observation, $L_{\gamma}$ to the bolometric luminosity of the source 
and $\omega_{\gamma}$ is the fraction of sources that are below the detection threshold and thus contribute to the IGRB. 
We integrate up to redshift of $z_{\textrm{max}}=4$, for $\Gamma \in [1.6, 2.4]$ and include sources with luminosities from $0.7 \times 10^{44}$ to $10^{50}$
erg/s. 

\begin{figure}
\begin{centering}
\hspace{-0.0cm}
\includegraphics[width=3.6in,angle=0]{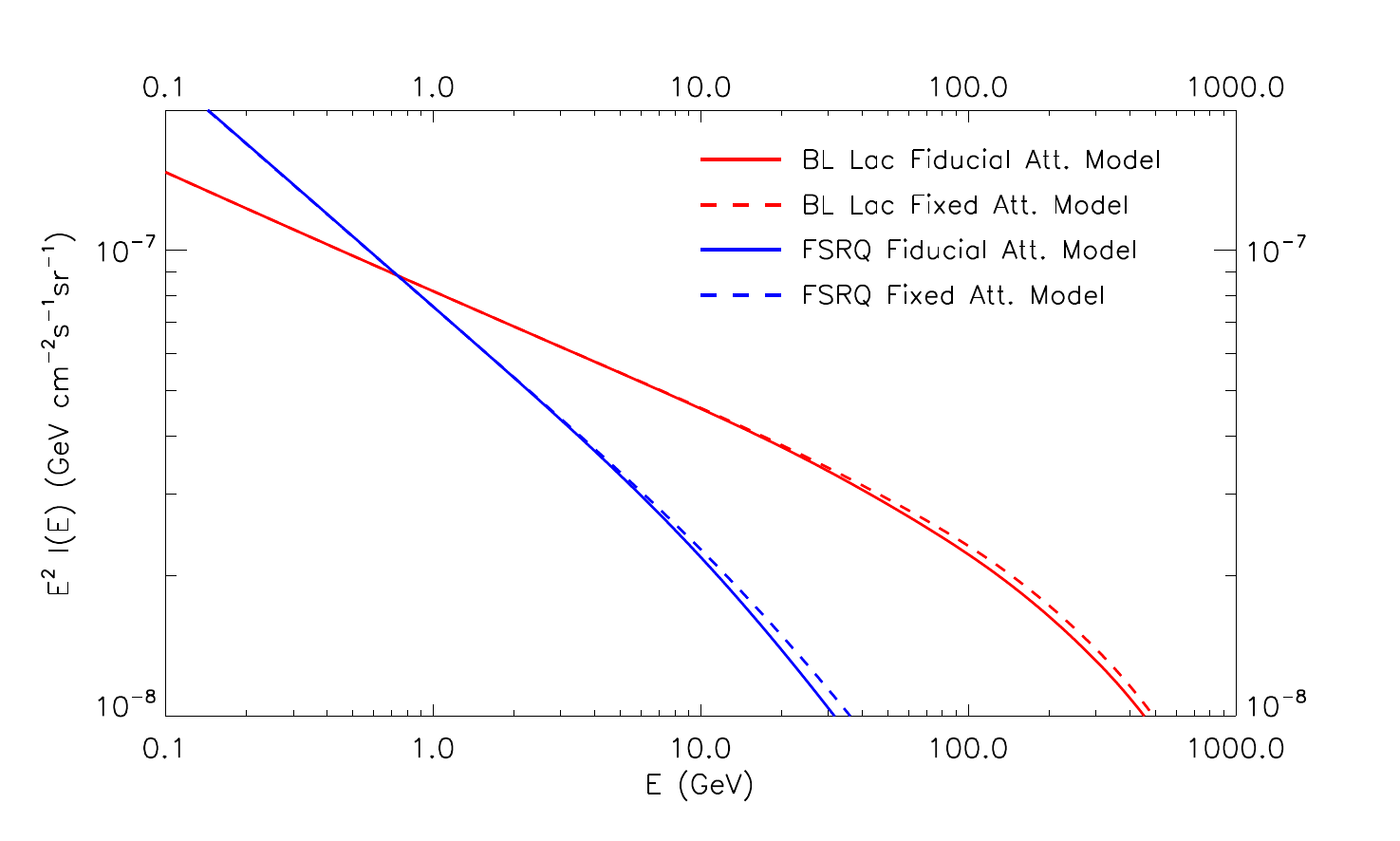}
\end{centering}
\vspace{-0.6cm}
\caption{
The contribution to the IGRB spectrum from unresolved BL Lac (red lines) and FSRQ (blue lines) sources.
The solid (dashed) lines assume the ``fiducial'' (``fixed'') attenuation model of Ref.~\cite{2012MNRAS.422.3189G}.
In fitting the IGRB spectrum from all sources contributing to it, the flux normalization and spectral index of each 
component are allowed to vary (see text for more details).}
\vspace{-0.7cm}
\label{fig:BLLac_FSRQ}
\end{figure}

In Fig.~\ref{fig:BLLac_FSRQ}, we show the contribution to the IGRB spectrum from BL Lacs and FSRQs.
We take $\mu^{*}_{\textrm{BL Lac}} = 2.3$ and $\mu^{*}_{\textrm{FSRQ}} = 2.5$ and assume for those lines 
that $\omega_{\gamma} =0.9$. We show results for both attenuation models that we test.  The normalization and spectral 
index of each component is allowed to vary once combining with the other astrophysical sources contributing to the IGRB.

\subsection{Modeling the contribution from Star-forming and Starburst Galaxies}
\label{subsec:Starforming}

There is only a small number of star-forming and star-burst galaxies without an AGN that have been identified as 
gamma-ray sources. However, such galaxies produce gamma rays via interactions of cosmic rays with their
respective interstellar media. To built a model on their redshift and luminosity distribution we rely on their 
infrared (IR) emission and correlate the two luminosities (in gamma rays and in infrared) via Ref.~\cite{2012ApJ...755..164A},
\begin{equation}
log_{10}\left( \frac{L_{\gamma}}{\textrm{erg s}^{-1}}\right) = \alpha \cdot log_{10} \left( \frac{L_{\textrm{IR}}}{10^{10} L_{\odot}}\right) + \beta.
\label{eq:LIR_to_Lgamma}
\end{equation}
Parameters $\alpha$ and $\beta$ are taken to be 1.17 and 39.28 respectively. 

The IR bolometric luminosity-function differs between classes of galaxies. Ref.~\cite{Gruppioni:2013jna},
using \textit{Herschel} observations, has provided us with the IR luminosity function  and its evolution
with redshift for different populations of galaxies. We use the information provided in Table 8 
of Ref.~\cite{Gruppioni:2013jna}, to evaluate the IR bolometric luminosity of three galaxy-populations;
regular star-forming (SF) galaxies, star-forming galaxies containing an AGN (SF-AGN) and star-burst (SB) galaxies.
Since the IR emission from a star-forming galaxy with an AGN, is dominated by the star-forming region and not the 
AGN itself, the gamma-ray emission that we evaluate from Eq.~\ref{eq:LIR_to_Lgamma}, is the one originating from
the start-forming region and not the AGN. 

The resulting gamma-ray luminosity function from the combined populations is,
\begin{eqnarray}
\Phi_{\gamma} (L_{\gamma}, z, f_{sb}) &=& (1 - f_{sb}) \cdot [\Phi_{\textrm{IR}}^{\textrm{SF}}(L_{\textrm{IR}}(L_{\gamma}),z,\{g^{\textrm{SF}}\}) \nonumber \\
&+& \Phi_{\textrm{IR}}^{\textrm{SF-AGN}}(L_{\textrm{IR}}(L_{\gamma}),z,\{g^{\textrm{SF-AGN}}\}) ] \nonumber \\
&+& f_{sb} \cdot \Phi_{\textrm{IR}}^{\textrm{SB}}(L_{\textrm{IR}}(L_{\gamma}),z,\{g^{\textrm{SB}}\}).
\label{eq:SF_GLF}
\end{eqnarray}
$f_{sb}$ is the fraction in the luminosity function that is contributed by the SB galaxies, which we take to be 0.5 for reference.
$L_{\textrm{IR}}$ is related to $L_{\gamma}$ based on Eq.~\ref{eq:LIR_to_Lgamma} and $\{g^{\textrm{SF}}\}$,
$\{g^{\textrm{SF-AGN}}\}$ and $\{g^{\textrm{SB}}\}$ parametrize the IR luminosity functions of SF, SF-AGN and SB galaxies,
$\Phi_{\textrm{IR}}^{\textrm{SF}}$, $\Phi_{\textrm{IR}}^{\textrm{SF-AGN}}$ and $\Phi_{\textrm{IR}}^{\textrm{SB}}$ respectively.
Full details on the latter are provided in Ref.~\cite{Gruppioni:2013jna}.

We take the spectrum of gamma rays 
produced \textit{on average} at these galaxies to be (before attenuation),
\begin{eqnarray}
\frac{dN}{dE} \propto && \left( \frac{E}{\textrm{1 GeV}} \right)^{-\gamma_{1}} \cdot \left(1 - exp\left[ \frac{-E}{E_{0}} \right] \right) \nonumber \\
&+&  \left( \frac{E}{\textrm{1 GeV}} \right)^{-\gamma_{2}} \cdot exp\left[ \frac{-E}{E_{0}} \right].
\label{eq:SFG_Spectrum}
\end{eqnarray}
We chose for $\gamma_{1}=2.2$, $\gamma_{2} = 1.95$ and $E_{0} = 0.3$ GeV, to parametrize the expected gamma-ray 
emission spectrum from the combination of $\pi^0$-decay, Bremsstrahlung emission and inverse Compton scattering processes 
in galaxies. As with the BL Lac and FSRQ objects, we allow in the fit process some deformation on the combined spectrum 
coming from star-forming and star-burst galaxies. 

\begin{figure}
\begin{centering}
\hspace{-0.0cm}
\includegraphics[width=3.6in,angle=0]{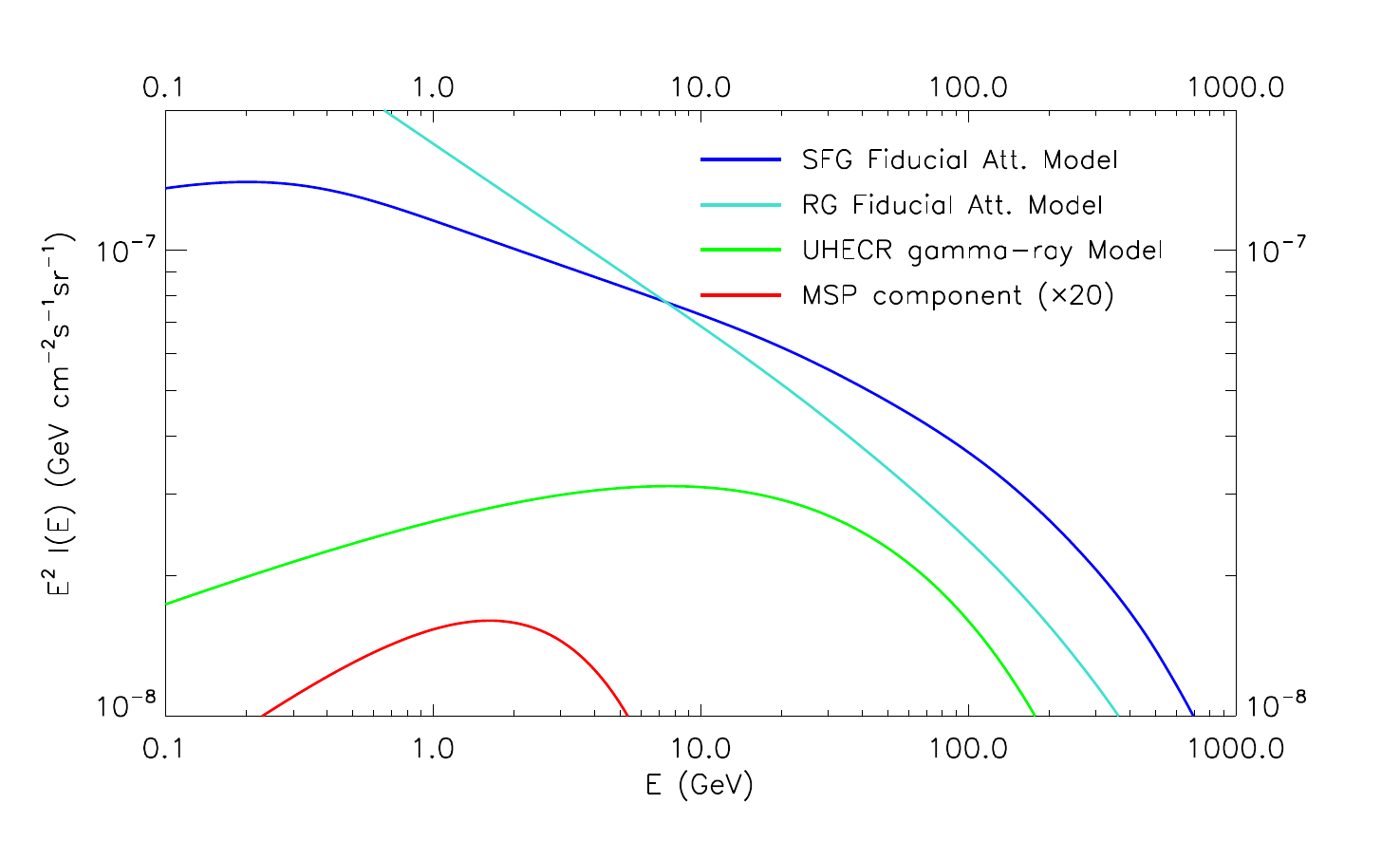}
\end{centering}
\vspace{-0.7cm}
\caption{
The contribution to the IGRB spectrum from unresolved star-forming and star-burst galaxies ``SFG'' (blue line), from radio 
galaxies ``RG'' (turquoise line), from the interaction of UHECRs with the intergalactic medium (green line) and from the galactic MSPs (red line).
We use the ``fiducial'' attenuation model of Ref.~\cite{2012MNRAS.422.3189G} for the ``SFG'' and ``RG'' components (see text for details).
Shown normalizations are chosen for all four components to be within the plotted intensity range, while in fitting to the IGRB spectrum the individual 
flux normalizations and spectral indices are allowed to vary.}
\vspace{-0.6cm}
\label{fig:SFG_RG_UHECR_MSP}
\end{figure}

The total intensity in gamma rays from star-forming and star-burst galaxies is evaluated in a similar manner as for BL Lac and 
FSRQ sources in Eq.~\ref{eq:I_AGN}, with the difference that there is no integration over a range of spectral indices $\Gamma$ 
and that we take $\omega_{\gamma}=0$ for these objects as a very small number of close-by SF galaxies have been identified 
as sources. We integrate to a maximum redshift of 5 and assume $L_{\gamma}$ in the range of $10^{37}-10^{41}$ erg/s. As with 
Sec.~\ref{subsec:BLLac-FSRQ}, we test both the ``fiducial'' and ``fixed'' attenuation models. In Fig.~\ref{fig:SFG_RG_UHECR_MSP},
we show the contribution of star-forming and star-burst galaxies to the IGRB intensity. 

\subsection{Modeling the contribution from Radio Galaxies}
\label{subsec:Radio}

Gamma-ray emission from radio galaxies originates dominantly from the AGN jets in them. 
Such galaxies have misaligned (to our line of sight) relativistic jets, with strong radio emission form 
formed lobes. They are classified as Fanaroff-Riley (FR) type I and II radio galaxies; with the former 
having edge-darkened lobes and the latter edge-brightened. FRI and FRII sources are the much more 
numerous misaligned BL Lac and FSRQ blazars, based on the AGN unification scheme \cite{Urry:1995mg}. 
Their contribution to the EGRB and the IGRB is expected to be dominant \cite{2011ApJ...733...66I, 
Stecker:2010di, Inoue:2011bm, DiMauro:2013xta, Hooper:2016gjy, Fukazawa:2022gwm}, even though only a small 
number of them have been detected in the gamma-ray catalogues \cite{Fermi-LAT:2019yla}. 

To model their redshift and luminosity distribution and spectral properties in gamma rays, we rely instead 
on their emission properties at radio waves. We use the correlation relation \cite{DiMauro:2013xta, Hooper:2016gjy, Stecker:2019ybn}, 
\begin{equation}
log_{10}\left( \frac{L_{\gamma, \, \textrm{RG}}}{\textrm{erg s}^{-1}}\right) = b \cdot log_{10} \left( \frac{L_{\textrm{5 GHz}}}{10^{40} \textrm{erg s}^{-1}}\right) + d.
\label{eq:L5GHz_to_Lgamma}
\end{equation}
The gamma-ray luminosity $L_{\gamma, \, \textrm{RG}}$ is evaluated in the energy range of 0.1-100 GeV and $L_{\textrm{5 GHz}}$ is the 
core luminosity at radio waves at 5 GHz. However, since there is a significant scatter between galaxies that correlation 
is more accurately described by the probability that a galaxy with $L_{\textrm{5 GHz}}$ has $L_{\gamma, \, \textrm{RG}}$.
\begin{eqnarray}
&&P(L_{\gamma, \, \textrm{RG}}, L_{\textrm{5 GHz}}) = \frac{1}{\sqrt{2 \pi \sigma_{\textrm{RG}}^2}} \\
&\times& exp \left[ - \frac{\left( log_{10}\left( \frac{L_{\gamma, \, \textrm{RG}}/(\textrm{1 erg s}^{-1})}{ (L_{\textrm{5 GHz}}/(10^{40} \textrm{ erg s}^{-1}))^{b_{\textrm{RG}}} }\right) -d_{\textrm{RG}} \right)^2}{2 \sigma_{\textrm{RG}}^2} \right]. \nonumber
\label{eq:Prop_Lgamma}
\end{eqnarray}
Following Ref.~\cite{Blanco:2021icw}, we take $b_{\textrm{RG}} = 0.78$, $d_{\textrm{RG}} = 40.78$ and a spread $\sigma_{\textrm{RG}} = 0.88$.

To evaluate a gamma-ray luminosity function $\Phi_{\gamma}$ from radio galaxies, we need first to model the radio luminosity 
of the cores of these galaxies at 5 GHz, $\Phi_{\textrm{RG c}}$. That radio luminosity function is \cite{Blanco:2021icw}, 
\begin{eqnarray}
&& \Phi_{\textrm{RG c}}(L_{\textrm{5 GHz}}, z) = \Phi_{\textrm{RG}}^{\star}(z)  \\
&\times& \left( \left( \frac{L_{\textrm{5 GHz}}}{L_{\textrm{5 GHz}}^{\star}(z)}\right)^{\beta_{\textrm{RG}}}
+  \left( \frac{L_{\textrm{5 GHz}}}{L_{\textrm{5 GHz}}^{\star}(z)}\right)^{\gamma_{\textrm{RG}}} \right)^{-1}. \nonumber 
\label{eq:Phi_5GHz}
\end{eqnarray}
With,
\begin{eqnarray}
\Phi_{\textrm{RG}}^{\star}(z) &=& e_{1}(z)\phi_{1, \textrm{RG}}, \\
L_{\textrm{5 GHz}}^{\star}(z)&=& \frac{L^{\star}}{e_{2}(z)}, \\
e_{1}(z) &=& \frac{(1 + z_{c})^{p_{1}} + (1 + z_{c})^{p_{2}}}{\left(\frac{1 + z_{c}}{1 + z}\right)^{p_{1}} + \left(\frac{1 + z_{c}}{1+z}\right)^{p_{2}}}, \\
e_{2}(z) &=& (1 + z)^{k_{1}}.
\label{eq:Phi_5GHz_specifics}
\end{eqnarray}
Following Ref.~\cite{2018ApJS..239...33Y}, we take $\phi_{1, \textrm{RG}} = 10^{-3.75} \; \textrm{Mpc}^{-3}$, $L^{\star} = 10^{28.59} \times (5 \times 10^{9}) \; \textrm{erg s}^{-1}$, $z_{c} = 0.893$, $p_{1} =2.085$, $p_{2} = -4.602$, $\beta_{\textrm{RG}} =0.139$ and $\gamma_{\textrm{RG}} = 0.878$.

The two luminosity functions are then related via,
\begin{eqnarray}
\Phi_{\gamma}(L_{\gamma}, z) = \frac{1}{b_{\textrm{RG}}} \int_{L_{\textrm{5 GHz}}^{\textrm{min}}}^{L_{\textrm{5 GHz}}^{\textrm{max}}}
&& \frac{d L_{\textrm{5 GHz}}}{L_{\textrm{5 GHz}}} \cdot \Phi_{\textrm{RG c}}(L_{\textrm{5 GHz}}, z) \nonumber \\
&\times& P(L_{\gamma}, L_{\textrm{5 GHz}}), 
\label{eq:Phi_gamma}
\end{eqnarray}
where we integrated between $L_{\textrm{5 GHz}}^{\textrm{min}} = 10^{37}  \textrm{erg s}^{-1}$ and 
$L_{\textrm{5 GHz}}^{\textrm{max}} = 10^{43}  \textrm{erg s}^{-1}$.

We take the spectrum of gamma rays 
produced on average at these galaxies to be (before attenuation),
\begin{equation}
\frac{dN}{dE} \propto \left( \frac{E}{\textrm{1 GeV}} \right)^{-\gamma_{\textrm{RG}}}.
\label{eq:RG_Spectrum}
\end{equation}
The index $\gamma_{RG}$, is taken to be 2.39.

The total intensity of gamma rays from radio galaxies is,
\begin{eqnarray}
I(E_{\gamma}) &=& \int_{0}^{z_{\textrm{max}}}dz 
\int_{L_{\gamma}^{\textrm{min}}}^{L_{\gamma}^{\textrm{max}}} \frac{dV}{dz}  \Phi_{\gamma} (L_{\gamma}, z) \cdot \frac{dN}{dE}(E_{\gamma}(1+z)) \nonumber \\
&\times& \frac{L_{\gamma}}{2 \pi (d_{L}(z))^{2}} \cdot exp \left[ - \tau(z, E_{\gamma}(1+z)) \right].
\label{eq:I_RG}
\end{eqnarray}
Again,  $E_{\gamma}$ is the gamma-ray energy at observation.
We integrate up to redshift of $z_{\textrm{max}}=5$, for luminosities from $10^{38}$ to $10^{42.5}$
erg/s and test both the ``fiducial'' and ``fixed'' attenuation models.  
In Fig.~\ref{fig:SFG_RG_UHECR_MSP},
we show a reference spectrum from radio galaxies to the IGRB intensity. 

\subsection{Modeling the contribution from Ultra-High-Energy Cosmic Rays}
\label{subsec:UHECRs}

Ultra-high energy cosmic-ray nuclei (including protons) scatter with the intergalactic infrared background and with the cosmic microwave background.
As a result their flux gets attenuated \cite{1966PhRvL..16..748G, Zatsepin:1966jv}, while also giving a flux of gamma rays.
These gamma rays are produced either directly from the resulting cascades or from high-energy electrons and positrons that 
have pair annihilation and inverse Compton scattering interactions. 
Like with the other extragalactic in nature sources of gamma rays 
the flux of gamma rays from UHECRs gets attenuated.
We take the spectrum of resulting gamma rays to be (including attenuation), 
\begin{equation}
\frac{dN}{dE} = A_{\textrm{UHE}} \left( \frac{E}{\textrm{1 GeV}} \right)^{-\gamma_{\textrm{UHE}}} exp\left[ -\left( \frac{E}{E_{\textrm{cut}}}\right)^{\beta_{\textrm{UHE}}} \right],
\label{eq:UHECR_to_gamma_ray_Spectrum}
\end{equation}
with a reference normalization $A_{\textrm{UHE}} = 3.0 \times 10^{-8}$ $\textrm{GeV}^{-1} \textrm{cm}^{-2} \textrm{s}^{-1} \textrm{sr}^{-1}$, $\gamma_{\textrm{UHE}} = 1.78$, $\beta_{\textrm{UHE}} = 0.54$ and $E_{\textrm{cut}} = 40$ GeV \cite{Cholis:2013ena}. 

We note that this spectrum depends on the redshift distribution of UHECR sources, their chemical composition, extragalactic magnetic fields and the infrared background. All these come with large uncertainties affecting the overall flux of gamma-rays from UHECRs. As a result, in our fits we allow for the 
gamma-ray flux of Eq.~\ref{eq:UHECR_to_gamma_ray_Spectrum} a large uncertainty in its normalization from the reference 
spectrum. In Fig.~\ref{fig:SFG_RG_UHECR_MSP},
we plot the contribution from UHECRs to the IGRB intensity for the spectral choices of Eq.~\ref{eq:UHECR_to_gamma_ray_Spectrum}.

\subsection{Modeling the contribution from Galactic Millisecond Pulsars}
\label{subsec:MSPs}

Millisecond pulsars are recycled neutron stars that have been spun up by accreting matter from their
companion star. Their spectrum can be derived from the \textit{Fermi}-LAT observations. 
Following Ref~\cite{Cholis:2013ena}, that relied on modeling the spectrum and luminosity properties of these
galactic sources from \cite{Cholis:2014noa, Cholis:2014lta}, we take the spectrum
of gamma-ray emission from unresolved MSPs to be, \cite{Cholis:2014noa},
\begin{equation}
\frac{dN}{dE} = A_{\textrm{MSP}} \left( \frac{E}{\textrm{1 GeV}} \right)^{-\gamma_{\textrm{MSP}}} exp\left[ -\left( \frac{E}{E_{\textrm{cut MSP}}}\right) \right],
\label{eq:MSP_Spectrum}
\end{equation} 
with $\gamma_{\textrm{MSP}} = 1.57$ and $E_{\textrm{cut MSP}} = 3.78$ GeV. 
In Fig.~\ref{fig:SFG_RG_UHECR_MSP}, we give the MSP spectrum for a reference value $A_{\textrm{MSP}} = 1.0 \times 10^{-9}$ $\textrm{GeV}^{-1} \textrm{cm}^{-2} \textrm{s}^{-1} \textrm{sr}^{-1}$, that is in agreement with the constraints from \textit{Fermi} data on those source's contribution to the gamma-ray 
 anisotropy measurement of Ref.~\cite{2011MNRAS.415.1074S} (see also \cite{Hooper:2013nhl}).

\subsection{Combining all known gamma-ray sources to the EGRB}
\label{subsec:Combination}

Before considering the possible contribution of dark matter, we fit the \textit{Fermi} IGRB spectrum to a combination
of the astrophysical sources described above. As we described earlier each component has a relevant freedom
in its normalization and its spectral index, $\Delta_{\gamma}$. Only for the UHECR-related component and for the MSPs do we fix their spectral shape to that described in Secs.~\ref{subsec:UHECRs},~\ref{subsec:MSPs} and Eqs.~\ref{eq:UHECR_to_gamma_ray_Spectrum} and~\ref{eq:MSP_Spectrum}.
We note that the MSP component is the least important flux component to the
IGRB.
In Table~\ref{tab:FitFreedom}, we give the relevant ranges on the normalization of each component from the
reference values described in the previous sections and the range of spectral index change $\Delta \gamma$. The ranges 
are chosen to account generously for the relevant modeling uncertainties. In Appedndix~\ref{app:background_scrict_fit}, we present a more constrained parameter scan used for the modeling of the IGRB backgrounds. Our dark matter limits are practically unaffected by the alternative choices in marginalizing over the background components normalizations and spectral indices.  

\begin{table}[t]
    \begin{tabular}{ccc}
         \hline
           Component &  Normalization Range & $\Delta{\gamma}$ \\
            \hline \hline
 g            BLLac &  [0.2, 1.2] & [-0.15, + 0.15] \\   
            SFRQ &  [0.5 1.5] & [-0.04, + 0.03] \\
            SF \& SB & [0.1, 1.0] & [-0.3, +0.3] \\
            RG &  [0.2, 1.5] & [-0.3, +0.2] \\
            UHECR &  [0.2, 5] & 0.0 \\     
            MSP &  [0.5, 1.5] & 0.0 \\
        \hline \hline 
        \end{tabular}
        \vspace{-0.3cm}
\caption{The freedom in normalization and spectral shape of each background astrophysical component when fitting to the IGRB spectrum.} 
\vspace{-0.5cm}
\label{tab:FitFreedom}
\end{table}

\begin{figure}
\begin{centering}
\hspace{-0.0cm}
\includegraphics[width=3.6in,angle=0]{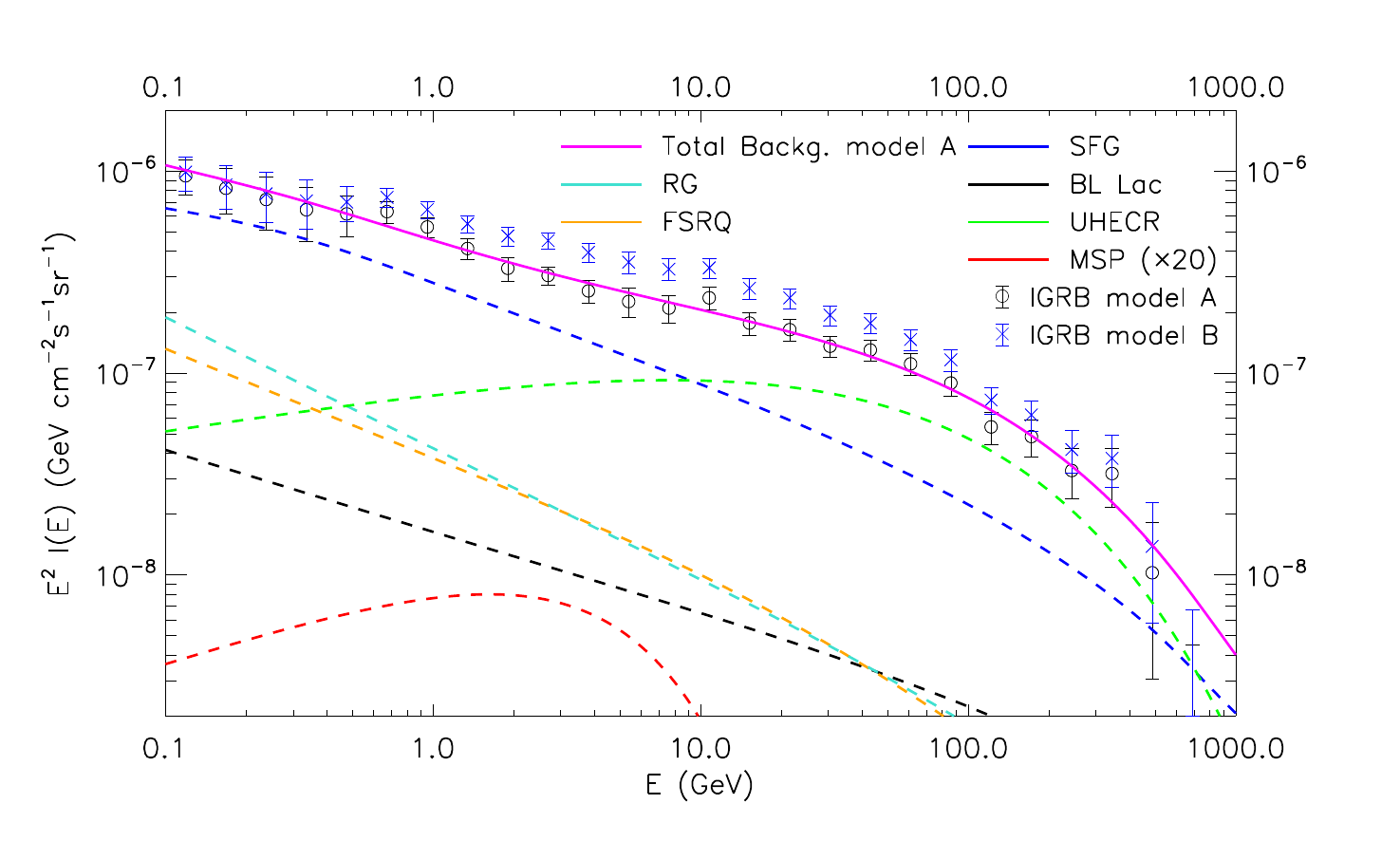}
\end{centering}
\vspace{-0.7cm}
\caption{
Best fit model to the IGRB spectrum from the combination of star-forming and star-burst galaxies ``SFG'' (blue dashed line), from radio 
galaxies ``RG'' (turquoise dashed line), from BL Lac sources (black dashed line), from FSRQ sources (orange dashed line), from UHECRs (green dashed line) and from the galactic MSPs (red dashed line $\times 20$). 
Each component is shown with the normalization coming from the best fit of $\chi^{2}$/dof=0.79, to the IGRB spectrum using model A from Ref.~\cite{Fermi-LAT:2014ryh} (black circles and errors). 
All components combined give us the total background emission ``Total Backg. model A'' (solid magenta line). We 
also provide in the blue ``x'' and errors for reference the IGRB spectrum using model B of Ref.~\cite{Fermi-LAT:2014ryh}.}
\vspace{-0.6cm}
\label{fig:BackModA}
\end{figure}

In Fig.~\ref{fig:BackModA}, we give the best-fit combination of all background astrophysical sources (magenta solid line) to the IGRB spectrum using the spectral model A of Ref.~\cite{Fermi-LAT:2014ryh}  (black circles and errors). We also show the contribution from each component (in dashed lines).
Radio galaxies, star-forming galaxies and the cascade-produced gamma rays from UHECRs represent the three major contributions to the IGRB. 
We use the ``fiducial'' attenuation model of Ref.~\cite{2012MNRAS.422.3189G} for the star-forming  and radio galaxy components.
We perform the same kind of analysis for the IGRB spectrum using models B and C of Ref.~\cite{Fermi-LAT:2014ryh}.
For model B the spectrum is given for reference by the blue ``x'' and errors of Fig.~\ref{fig:BackModA}. 
The IGRB spectrum of model C, is not shown in that figure, as it significantly overlaps with model A.

As can be seen from Fig.~\ref{fig:BackModA}, given the freedom we allow in both the individual components normalizations and spectral index change in our fit, some components can become highly degenerate. 
The most typical example of sources that show degeneracy are FSRQs and Radio galaxies. Also, some amount of degeneracy is seen between star-forming galaxies and BL Lacs.
These degeneracies allow for a good fit to be achieved  by the combination of the background components. In this case we get a $\chi^{2}$/dof = 0.79.
As we aim to probe for a potential dark matter component in the IGRB that would allow for a better fit to the IGRB spectrum, our approach is conservative.

\subsection{Statistical analysis}
\label{subsec:analysis}

When we fit the astrophysical/background models to the IGRB spectrum, we minimize the $\chi^2$
loss with respect to the parameters listed in Table~\ref{tab:FitFreedom}. An example of such a fit is shown in Fig.~\ref{fig:BackModA}.
When we add an additional contribution from annihilating dark matter to the IGRB spectrum, an extra parameter enters our fit, which is the dark matter annihilation cross section.
This parameter acts as a normalization factor for the dark matter flux and scales the flux from its nominal value, which corresponds to an annihilation cross section (times velocity, thermally averaged) of $\langle \sigma v \rangle = 3.0 \times 10^{-26} \ \textrm{cm}^3/\textrm{s}$. Such fits are shown in our results section.

For our minimization process, we use \path{iminuit} \cite{iminuit,James:1975dr}.
The minimization proceeds as follows: For a given astrophysical background, we first minimize the $\chi^2$ loss with respect to only the astrophysical background parameters (i.e., the dark matter normalization is fixed to zero).
Next, we include a dark matter contribution with a specified dark matter mass, annihilation channel, and assumption for the halo mass function. We then perform the fit again to determine the best-fit values of all parameters, including the most important one: the dark matter annihilation cross section.

To establish upper limits on the dark matter annihilation cross section, represented by $\langle \sigma v \rangle$, we use a likelihood ratio test. The null hypothesis posits that there is no dark matter contribution, meaning that the spectrum is solely explained by the astrophysical background (i.e., $\langle \sigma v \rangle = 0$). In contrast, the alternative hypothesis introduces an additional component from dark matter annihilation, implying a non-zero annihilation cross section $\langle \sigma v \rangle$.

We apply Wilks' theorem \cite{10.1214/aoms/1177732360} to analyze this, using the test statistic $LR = -2\log{\Lambda}$, which is the difference in $\chi^2$ between the null hypothesis (background-only) and the alternative hypothesis (background plus dark matter). This test statistic follows a $\chi^2_{\nu}$ distribution, where $\nu$ is the number of additional fitting parameters in the alternative model compared to the null model. In our case, $\nu = 1$.

However, since the null hypothesis with $\langle \sigma v \rangle = 0$ is at the boundary of the parameter space, the p-value derived from this approach would be incorrect. To correct for this, we use Chernoff’s theorem \cite{10.1214/aoms/1177728725}, which shows that the $LR$ follows a distribution of $\frac{1}{2} \delta(x) + \frac{1}{2}\chi^2$, known as a half chi-square distribution, with one degree of freedom \cite{Conrad:2014nna}. This adjustment effectively reduces the p-value by half compared to the naive estimate.

In line with standard practices in the field, we determine 95\% upper limits on $\langle \sigma v \rangle$ for each astrophysical background by fixing the annihilation channel and dark matter mass. This process involves scanning through different values of $\langle \sigma v \rangle$, calculating the $\chi^2$ profile, and identifying the value of $\langle \sigma v \rangle$ where $\chi^2_{\mathrm{DM}} = \chi^2_{\mathrm{DM=0}} + 2.71$. This corresponds to the 95\% upper limit for a half chi-square distribution with one degree of freedom. At each point in this scan, the remaining background nuisance parameters are optimized to minimize the $\chi^2$ value.

This procedure is repeated for multiple dark matter masses.
The result is a 2D grid of dark matter mass points versus annihilation cross section, where we draw a 95\% upper limit line on the dark matter annihilation cross section as a function of mass.
This entire process is repeated for multiple dark matter annihilation channels and various assumptions of the halo mass function. Additionally, we use data from three different IGRB models (A, B, and C), so the entire analysis is repeated three times in total.

\section{The contribution of Dark Matter}
\label{sec:DM}

Dark matter annihilations may prominently contribute to the IGRB. 
There are two distinct components of dark matter contribution to the IGRB. 
One is the emission from extragalactic dark matter halos, from galaxy-cluster sized $ \sim 10^{15} M_{\odot}$
halos to as small in mass as $10^{3} M_{\odot}$ halos. The other is the emission from the Milky Way halo
and its substructure at high galactic latitudes. While not properly isotropic in nature as there
is a gradient of the main dark matter halo profile, that latter component is still contributing to the IGRB given the 
manner in which the \textit{Fermi} collaboration evaluates the IGRB spectrum.  Moreover, since dark matter
substructure survives at large galactocentric radii, the total contribution from the Milky Way halo to the IGRB has 
a smaller gradient with galactic latitude than the one expected by the purely Navarro-Frenk-White (NFW) 
profile. We describe the contribution of each of these two dark matter components in the following sections. 
We test models of dark matter annihilation to $\tau^{+} \tau^{-}$ leptons and to $b\bar{b}$ quarks 
with masses from 5 GeV and 6 GeV respectively and up to 5 TeV.
We also test dark matter annihilating to $W^{+} W^{-}$ (and $Z Z$) bosons with masses from 90 (100) GeV to 5 TeV. 
In all cases we include only the prompt dark matter gamma-ray emission which is by far more dominant to the 
gamma-ray emission due to inverse Compton scattering events or Bremsstrahlung emission.

\subsection{The emission from extragalactic dark matter halos}
\label{subsec:Extragalactic_DM}

The mean intensity of gamma rays of observed energy $E$ from dark matter annihilation due to overdensities 
$\delta = (\rho - \langle \rho \rangle)/ \langle \rho \rangle$ is given by \cite{Ando:2013ff},
\begin{equation}
I(E) = \int_{0}^{z_{\textrm{max}}} dz \,  \frac{c}{H(z)} W( (1+z)E, z) \cdot \langle \delta^{2} \rangle (z). 
\label{eq:DMannihilation_Intensity}
\end{equation} 
$H(z) = H_{0} \sqrt{  \Omega_{\Lambda} + \Omega_{k}(1+z)^2 + \Omega_{m} (1+z)^{3} + \Omega_{r}(1+z)^4 }$ 
with $H_{0} = 100 h$ the values for the Hubble expansion scaling factor $h$, and the dark energy density parameter $\Omega_{\Lambda}$, 
matter density parameter $\Omega_{m}$, radiation density parameter $\Omega_{r}$ and curvature parameter 
$\Omega_{k}$, taken from the \textit{Planck}-Collaboration results \cite{Planck:2018vyg}.

The window function $W(E, z)$ is,
\begin{equation}
W(E, z) = \frac{\langle \sigma v \rangle}{8 \pi} \left( \frac{\Omega_{dm} \rho_{c}}{m_{\chi}} \right)^{2} \, \frac{dN_{\gamma}}{dE} 
\, exp \left[ - \tau(z, E) \right],
\label{eq:Window_function}
\end{equation} 
with $\langle \sigma v \rangle$ the velocity-averaged annihilation cross-section of a dark matter 
particle of mass $m_{\chi}$. $\Omega_{dm}$ is the current dark matter density parameter and
$\rho_{c}$ the present critical density. $dN_{\gamma}/dE$ is the differential spectrum 
of produced gamma rays per annihilation event. We evaluate that spectrum from the prompt 
gamma-ray emission of each particle of mass $m_{\chi}$ and annihilation channel 
$\chi \chi \longrightarrow \tau^{+} \tau^{-}$ or $b\bar{b}$ or $W^{+} W^{-}$ or $Z Z$. For these 
spectra we use Ref.~\cite{Cirelli:2010xx}, that includes electroweak corrections relevant for the 
higher mass dark matter particles. 

The variance of the overdensities at redshift $z$ is \cite{Ando:2013ff},
\begin{eqnarray}
\langle \delta \rangle^{2} (z) = &&\left( \frac{1}{\Omega_{dm} \rho_{c}} \right)^{2} \int_{M_{\textrm{min}}}^{M_{\textrm{max}}} dM \, \frac{dn}{dM}(M,z) \nonumber \\
&\times&[1 + b_{\textrm{sh}}(M)] \int dV \rho_{\textrm{host}}^{2}(r,M).
\label{eq:deltaSq}
\end{eqnarray} 
$M$ is the halo's virial mass; while $dn/dM (M,z)$ the halo mass function (HMF) at a given redshift.
For the halo mass function we use the tabulated results provided by the \texttt{COLOSSUS}, toolkit \cite{2018ApJS..239...35D, COLOSSUSwebsite}, 
for halos with masses of $10^{3} - 10^{15}$ $M_{\odot}$ and redshifts up to $z=10$.
\texttt{COLOSSUS} provides us the possibility to test alternative HMFs. 
We have tested a long list of provided HMF tabulated results from the \texttt{COLOSSUS} toolkit.  
To envelop the impact of the uncertainties of the HMF on our results, we picked five alternate 
HMFs from \cite{2008ApJ...688..709T} ``Tinker et al. 08'',  \cite{2011MNRAS.410.1911C} ``Coutrin et al. 11'', 
\cite{2013MNRAS.433.1230W} ``Watson et al. 13'', \cite{2017MNRAS.469.4157C} ``Comparat et al. 17'' and from 
\cite{2021A&A...652A.155S} ``Seppi et al. 20''. 

The parameter $b_{\textrm{sh}}(M)$ of Eq.~\ref{eq:deltaSq}, is the boost factor to the annihilation signal due to substructures of a halo 
of mass $M$. This boost factor depends only very weakly on redshift. We follow Ref.~\cite{Gao:2011rf}, 
$b_{\textrm{sh}}(M) = 110 (M_{200}(M,z)/(10^{12} M_{\odot}))^{0.39}$, with $M_{200}$ the enclosed 
halo mass up to a radius at which the dark matter density is 200 times the critical one. For 
$M_{200}(M,z)$, we take the results of Ref.~\cite{Hu:2002we} (Appendix C), where, 
\begin{eqnarray}
M_{200}(M,z) &=& M (1 + g(c(M,z),y(M))), \textrm{\; with}\\
g(x,y) &=&  -3 (x+y) [ x -xy + (1+x)(x+y)ln(1+x) \nonumber \\
&& - (1+x)(x+y) ln(1 + x/y)] \nonumber \\
&& \times [3(1+x)(x+y)^{2}ln(1+x) \nonumber \\
&& - x(x+4x^{2} + 6xy +3y^{2})]^{-1}, \\
c(M,z) &=& 7.85 \cdot (1+z)^{-0.71} \left(\frac{M}{2 \times 10^{12} \, M_{\odot}} \right)^{-0.081}\\
\textrm{and} && \nonumber \\
y(M) &=& \left(\frac{\Delta_{h}}{\Delta_{v}}\right)^{1/3}. 
\label{eq:M_200_etc}
\end{eqnarray} 
For the concentration parameter $c(M,z)$ we use the parameterization of Ref.~\cite{Duffy:2008pz}, and test to alternative parameterization
of \cite{Hu:2002we}. 
These choices change our dark matter annihilation results by $\sim 1\%$. 
$y(M)$ is evaluated from the ratio of halo over-density 
$\Delta_{h}$ to virial over-density $\Delta_{v}$ (see Ref.~\cite{Hu:2002we} for more details). The exact assumptions on $y(M)$ can 
affect our results only at the $O(10^{-2})$ level.

Finally, the integral of the dark matter density squared can be analytically expressed for the NFW 
profile as \cite{Ando:2013ff, Moline:2016fdo},
 \begin{eqnarray}
\int dV \rho_{\textrm{host}}^{2}(r,M) &=& \frac{1}{9}\rho_{c}(z) \, \Delta(z) \cdot M \cdot c(M,z)^{3} \\
&\times& \frac{\left[1 - (1+ c(M,z))^{-3}\right]}{\left[ ln(1 + c(M,z)) - \frac{c(M,z)}{1+c(M,z)} \right]^{2}}.  \nonumber
\label{eq:deltaSq_2}
\end{eqnarray} 

\begin{figure}
\begin{centering}
\hspace{-0.0cm}
\includegraphics[width=3.6in,angle=0]{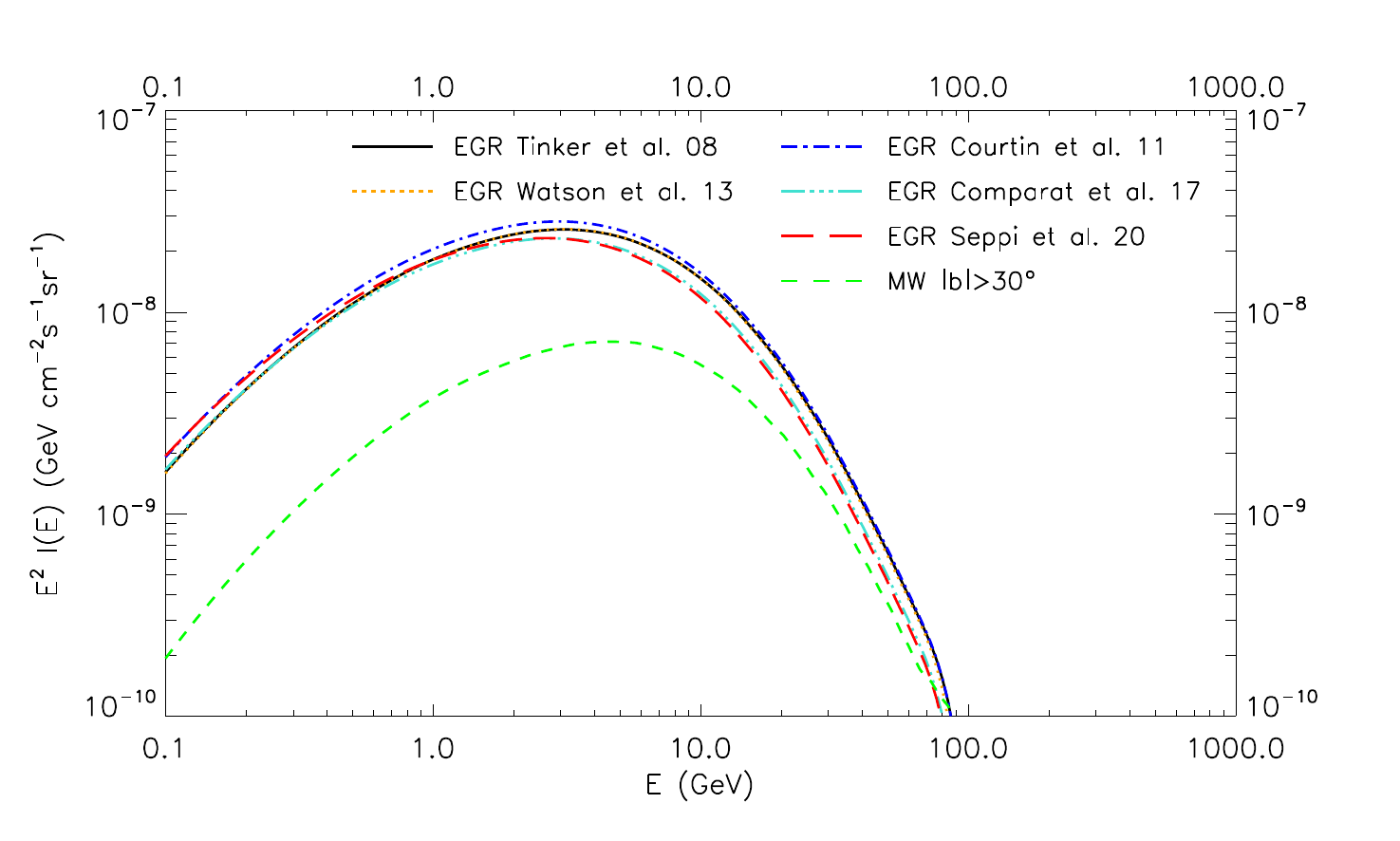}
\end{centering}
\vspace{-0.7cm}
\caption{
The dark matter contribution to the IGRB. We assume a 100 GeV particle annihilating to $b\bar{b}$ quarks with a cross-section of $\sigma v = 3.0\times 10^{-26} \textrm{cm}^3/\textrm{s}$. We show the dominant extragalactic dark matter component ``EGR'' for five choices of halo mass function and the subdominant Milky Way (MW) contribution evaluated at high latitudes (see text for details).}
\vspace{-0.6cm}
\label{fig:DMcomponents}
\end{figure}

In Fig.~\ref{fig:DMcomponents}, we show the contribution to the IGRB from dark matter annihilations 
taking place in galaxies other than ours. We refer to that as the extragalactic dark matter component (EGR). 
We show results for the five alternative HMFs.  We use the case of a $m_{\chi} = 100$ GeV mass particle 
annihilating to $b\bar{b}$ quarks with a cross-section of $\sigma v = 3.0\times 10^{-26} \textrm{cm}^3/\textrm{s}$.
The extragalactic dark matter component is the dominant part of the dark matter annihilation contribution to the IGRB.
The alternative HMF parameterizations give somewhat different spectral shapes due to differences in the redshift 
evolution of dark matter halos. Between them, a difference of up to $30\%$ in the expected gamma-ray flux is predicted.

\subsection{The contribution from the Milky Way halo to the Isotropic gamma-ray background}
\label{subsec:Galactic_DM}

Given the nature of the IGRB, dark matter annihilations in the Milky Way halo, occurring at high latitudes can contribute to its observed spectrum.
The intensity from these dark matter annihilations is given by,
\begin{equation}
I_{MW}(E) = \frac{\langle \sigma v \rangle}{2 m_\chi^2} \frac{dN_\gamma}{dE}  \frac1{\Omega_{hl}}  \int_{V_*} dV ~ \frac{ \rho^2(s,b,\ell)}{4\pi s^2}. 
\label{eq:MWcontr}
\end{equation} 
The $s$ parameter is the distance from the center of the halo, $\ell$ and $b$ are the galactic longitude and latitude respectively. $\Omega_{hl}$ is the solid angle observed at high latitudes. We take the dark matter to follow an NFW profile, with a scale radius $r_s=21.5$ kpc, a virial radius $r_{\rm vir}=258$ kpc. The local dark matter density at $r=8.5$ kpc is taken to 
$\rho_{\odot} = 0.3$ GeV cm$^{-3}$ in agreement with \cite{Iocco:2011jz, Bovy:2012tw, Catena:2009mf, Salucci:2010qr}. 
This gives a virial mass of the Milky Way of $M_{\rm vir}=1.6\times 10^{12}\, M_{\odot}$ consistent with the current estimates \cite{Klypin:2001xu}. 
We average the gamma-ray flux from latitudes of $|b| > 30^{\circ}$.
We ignore the contribution of Milky Way dark matter substructures, as at least for the most massive end, those are identified as separate objects, 
the dwarf galaxies orbiting the main Milky Way halo. This provides a conservative estimate of the dark matter annihilations in the Milky Way halo to the IGRB.
For a dark matter particle of $m_{\chi} = 100$ GeV, with an annihilation cross-section of $\sigma v = 3.0\times 10^{-26} \textrm{cm}^3/\textrm{s}$ to $b\bar{b}$ 
quarks we give our calculated gamma-ray flux in Fig.~\ref{fig:DMcomponents} (green dashed line). The annihilations in the Milky Way halo provide a 
contribution to the IGRB that is about a factor of 10 suppressed to the extragalactic dark matter contribution at low energies compared to the dark matter mass
$m_{\chi}$. At higher gamma-ray energies the ratio between the two dark matter components is closer to 1 as the Milky Way halo component is not redshifted 
nor attenuated. 

\section{Upper limits on the dark matter
annihilation cross-section}
\label{sec:DMlimits}

\begin{figure}[h]
\begin{centering}
\hspace{-0.0cm}
\includegraphics[width=3.6in,angle=0]{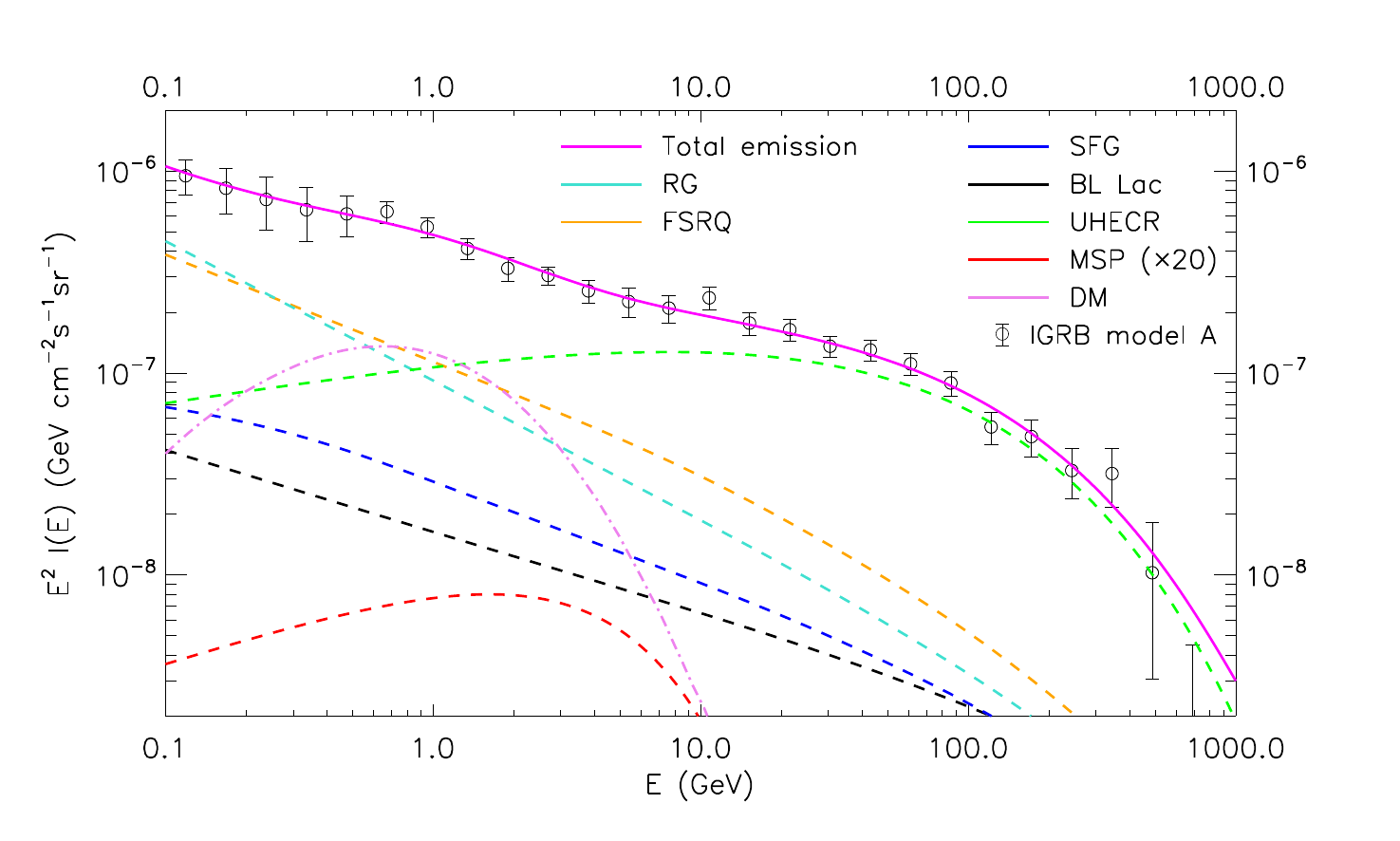}\\
\includegraphics[width=3.6in,angle=0]{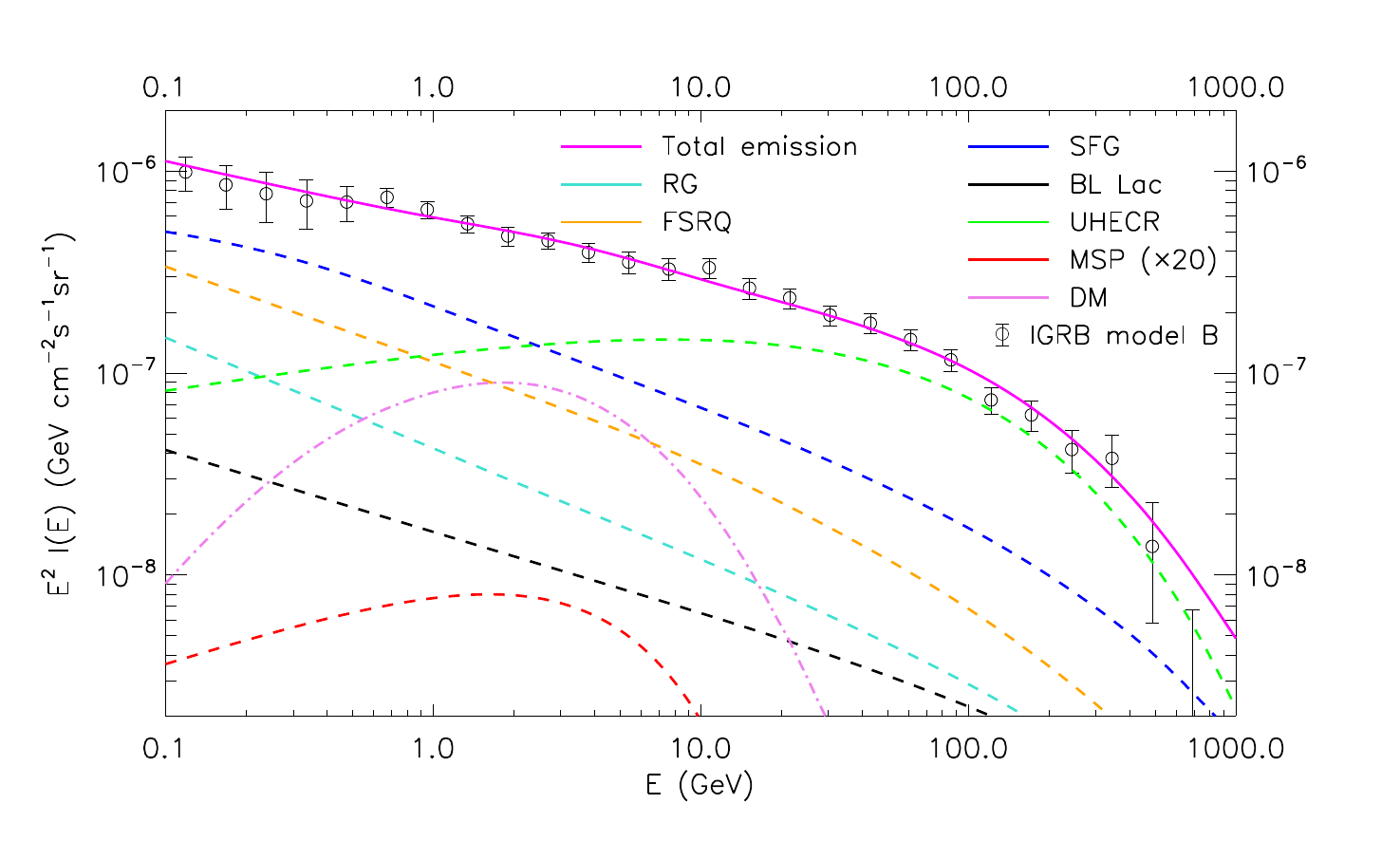}\\
\includegraphics[width=3.6in,angle=0]{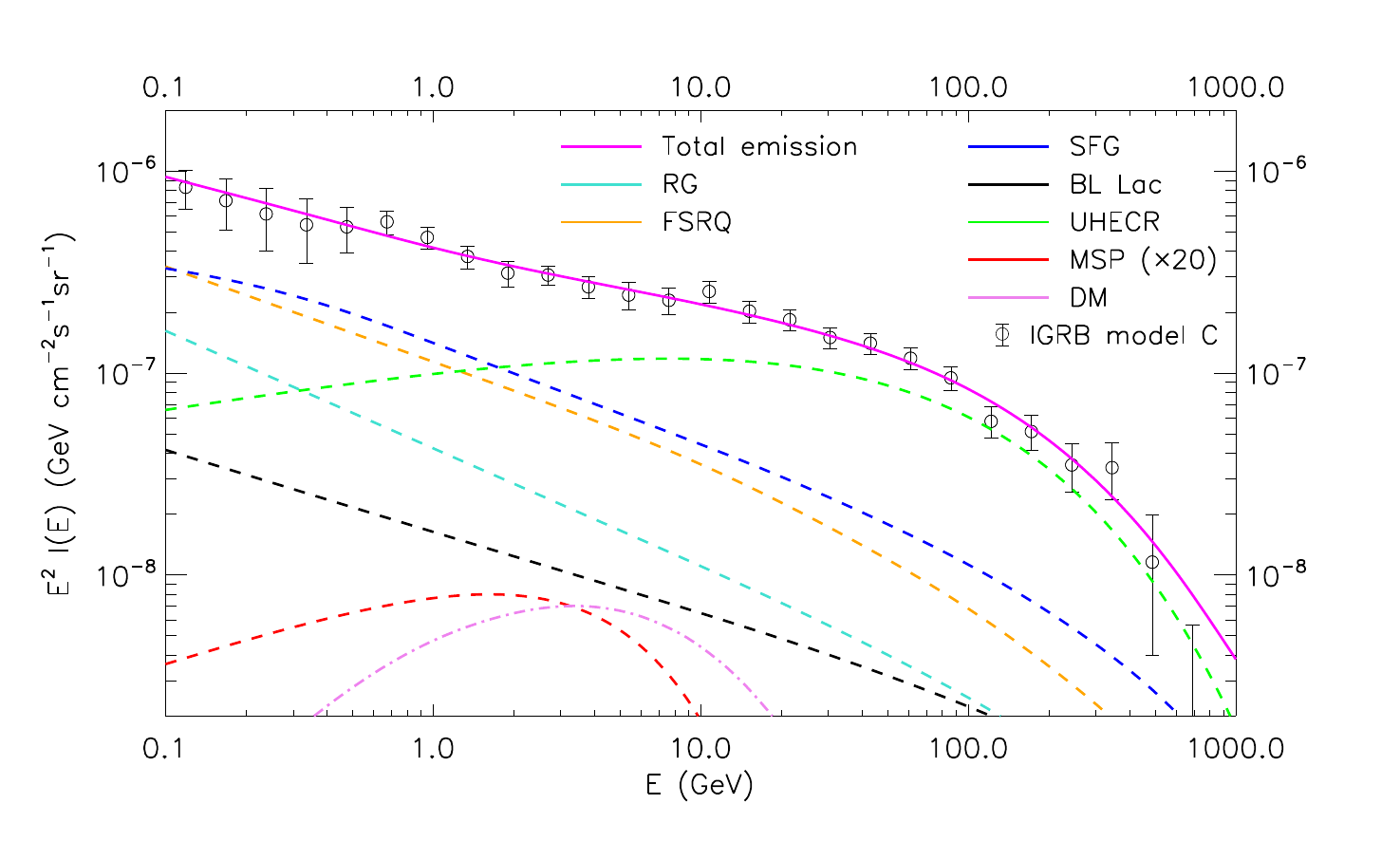}
\end{centering}
\vspace{-0.7cm}
\caption{
The contribution of all components to the IGRB. 
The top panel shows a fit to the IGRB spectrum A with a dark matter ``DM'' component (dashed-dotted violet line), assuming a 15 GeV particle annihilating to $b\bar{b}$ quarks, with a cross-section of $\sigma v = 1.6\times 10^{-26} \textrm{cm}^3/\textrm{s}$ (best fit value); leading a $\chi^{2}$/dof = 0.70. In the middle panel, we show a fit to the IGRB spectrum B, for a 50 GeV particle annihilating to $b\bar{b}$, with $\sigma v = 4.0\times 10^{-26} \textrm{cm}^3/\textrm{s}$; resulting in a $\chi^{2}$/dof = 0.75.
Finally, in the bottom panel for IGRB model C, we show the fit for a 100 GeV particle, with $\sigma v = 0.65\times 10^{-26} \textrm{cm}^3/\textrm{s}$; giving a $\chi^{2}$/dof = 0.72.}
\vspace{-0.6cm}
\label{fig:BestFitExamples}
\end{figure}

In Fig.~\ref{fig:BestFitExamples}, we show three examples of a fit to the IGRB data. In these three cases, we have found a positive contribution for a dark matter component. We show one result for each of the three IGRB spectral models (A, B and C of Ref.~\cite{Fermi-LAT:2014ryh}). For simplicity we use the same annihilation channel to $b\bar{b}$ quarks and the same HMF from Ref.~\cite{2008ApJ...688..709T} (Tinker08) for the calculation of the extragalactic dark matter contribution. 
The dark matter component ``DM'' in these figures is drown with a violet dashed-dotted line while the other background astrophysical components are shown with the same color choices as in Fig.~\ref{fig:BackModA}. 
Notice that dark matter gives a distinctively different spectrum than the conventional astrophysical populations. 
Thus, its contribution to the IGRB spectrum can be reliably probed. 
Millisecond pulsars also give a gamma-ray spectrum that is dissimilar to the extragalactic sources. However, their contribution to the gamma-ray emission at high latitudes has been strongly constrained as we discuss in Section~\ref{subsec:MSPs}.

\begin{figure*}[h]
\begin{centering}
\hspace{-0.2cm}
\includegraphics[width=3.4in,angle=0]{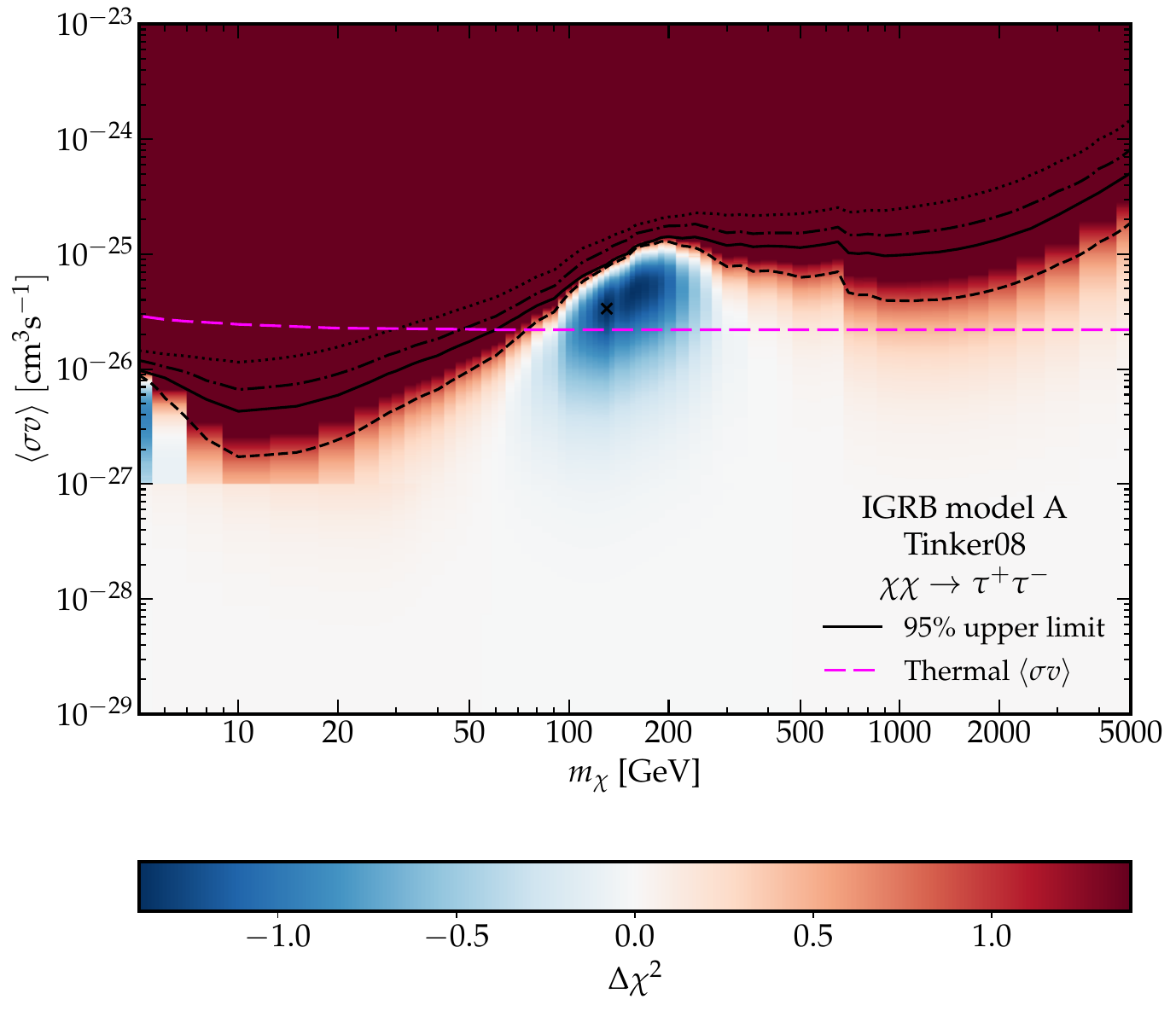}
\hspace{0.5cm}
\includegraphics[width=3.4in,angle=0]{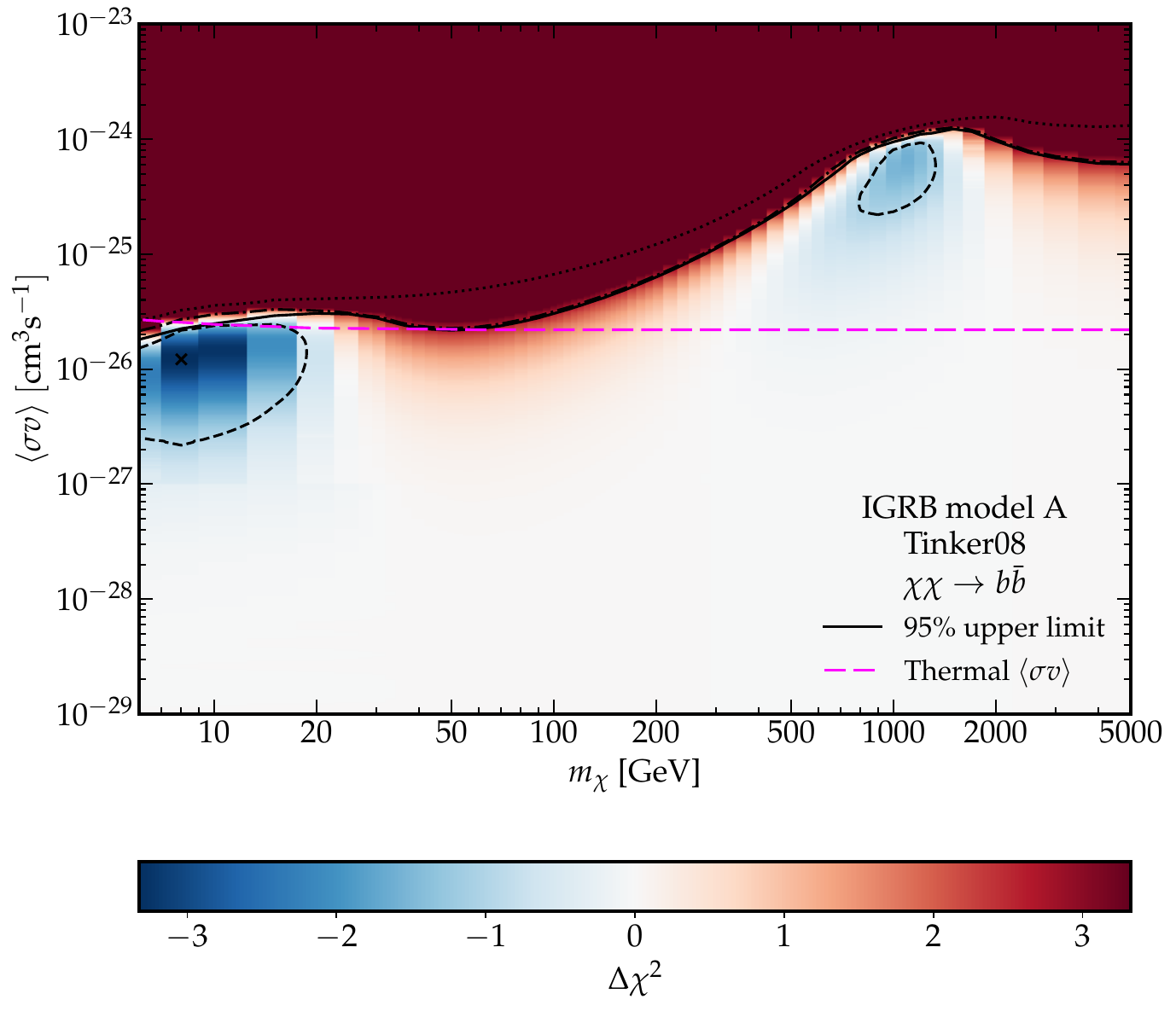}\\
\hspace{-0.2cm}
\includegraphics[width=3.4in,angle=0]{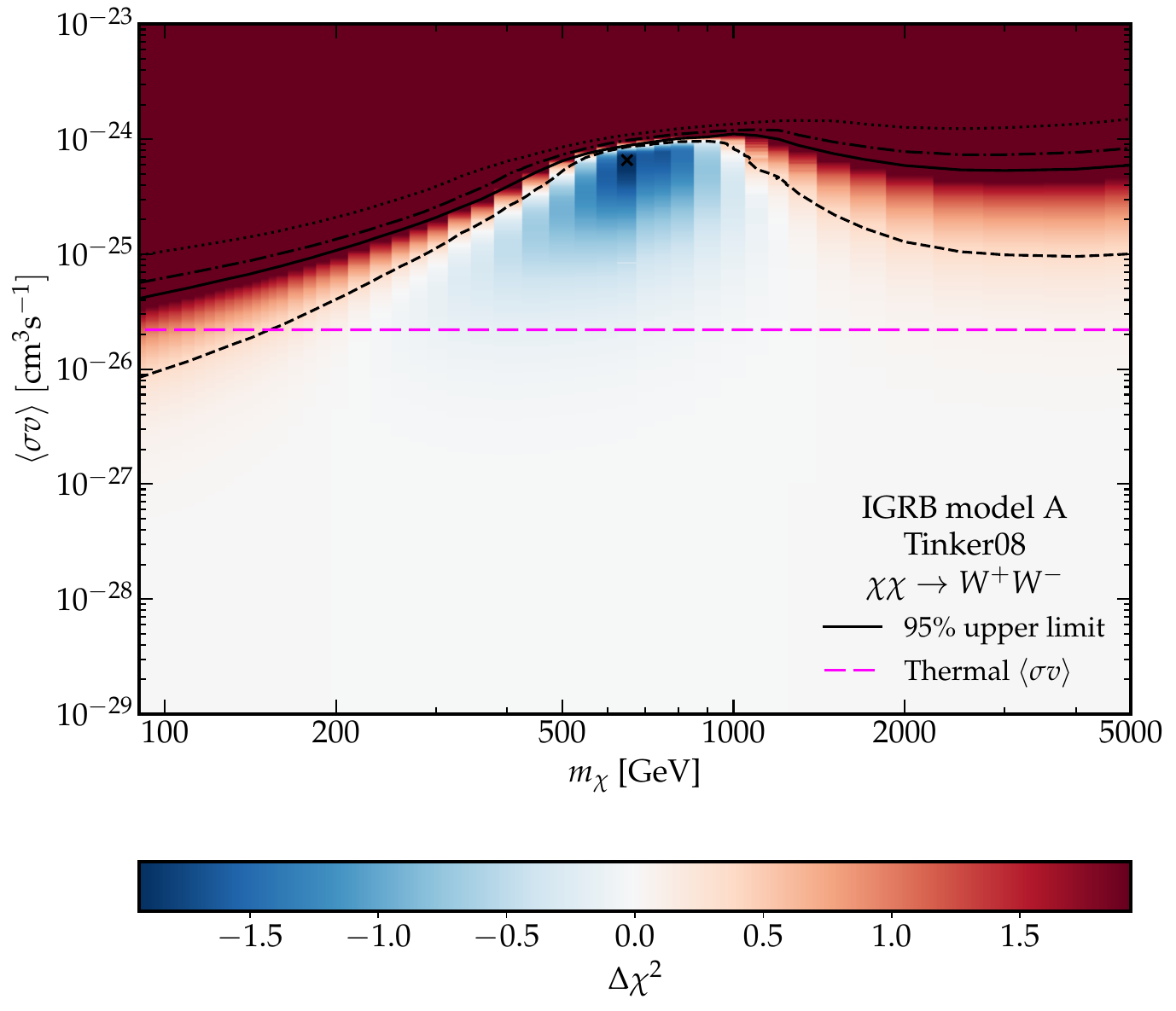}
\hspace{0.5cm}
\includegraphics[width=3.4in,angle=0]{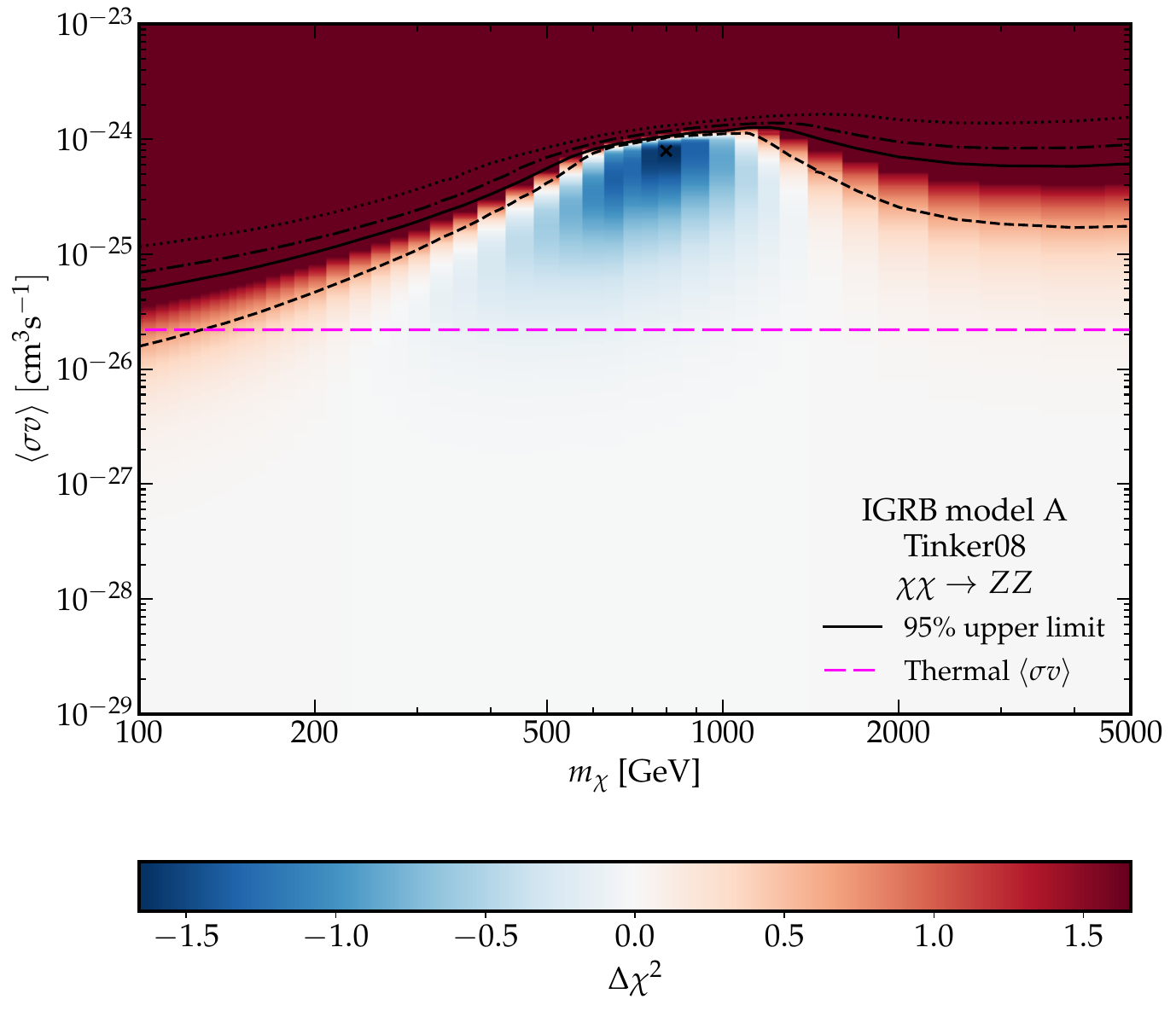}
\end{centering}
\vspace{-0.7cm}
\caption{
Using the IGRB model A spectrum, we show our fit results on the existence of a dark matter component. 
For every given mass, annihilation channel and cross-section $\langle \sigma v \rangle$, we evaluate the $\chi^{2}$ and compare its value to the one derived using only the background components. 
Blue regions show regions of the parameter space where the is preference for a dark matter contribution with the given assumptions, i.e. $\Delta \chi^{2} <0$ and red regions ($\Delta \chi^{2} >0$), where there is a penalty (see relevant color bar for the values of $\Delta \chi^{2}$). 
The magenta dashed line gives the thermal relic annihilation cross-section. The black ``x'' gives the best fit dark matter parameters for a given annihilation channel. 
The black dashed (dotted-dashed and dotted) lines give the region of dark matter parameter space that is within $1-\sigma$  ($2-\sigma$ and $3-\sigma$), from the best fit assumptions of that annihilation channel. 
The black solid line gives the $95 \%$ upper limit on the dark matter annihilation cross-section (see text for details).
The top left panel shows our results for the $\chi \chi \rightarrow \tau^{+} \tau^{-}$ annihilation channel. The top right panel for the $\chi \chi \rightarrow b\bar{b}$ channel. 
The bottom left panel for the $\chi \chi \rightarrow W^{+} W^{-}$ channel and the bottom right panel for $\chi \chi \rightarrow Z Z$.}
\vspace{-0.6cm}
\label{fig:DMlimits}
\end{figure*}

In Fig.~\ref{fig:DMlimits}, we show our fit results to a possible dark matter contribution. For any given combination of dark matter mass $m_{\chi}$, annihilation cross-section $\sigma v$ and channel (alternative panels), we evaluate the dark matter contribution to the IGRB spectrum. 
We then marginalize (within the quoted freedom of Table~\ref{tab:FitFreedom}) over the background components normalizations and spectral indices, to calculate a $\chi^{2}$ for that combination of dark matter parameters. 
We compare this $\chi^{2}$ value to the one derived using only the background components. 
A $\Delta \chi^{2}<0$ (blue regions) indicates a preference for a contribution to the IGRB flux from a dark matter particle with the given combination of mass, annihilation cross-section and channel. 
Instead, $\Delta \chi^{2}>0$ values (red regions), give the statistical penalty for adding a flux component from the relevant dark matter particle. We use the IGRB model A in this case. In Appendix~\ref{app:IGRB_B_and_C}, we present the equivalent results when using IGRB models B and C.

For some annihilation channels we find a small, less than $2\sigma$ preference for a thermal relic dark matter component.  
In fact, a dark matter particle that could explain the gamma-ray galactic center excess emission \cite{Hooper:2010mq, Vitale:2009hr, Abazajian:2010zy, Hooper:2011ti, Hooper:2013rwa, Gordon:2013vta,  Daylan:2014rsa, Calore:2014xka, Zhou:2014lva, Calore:2014nla, Agrawal:2014oha, Berlin:2015wwa, TheFermi-LAT:2015kwa, DiMauro:2021raz, Cholis:2021rpp, Zhong:2024vyi} and/or the cosmic-ray antiproton excess \cite{Cuoco:2016eej, Cui:2016ppb, Cuoco:2019kuu, Cholis:2019ejx}, is consistent with the $95 \%$ upper limits shown in Fig.~\ref{fig:DMlimits} (and also in Appendix~\ref{app:IGRB_B_and_C}). 

\begin{figure*}
\begin{centering}
\hspace{-0.2cm}
\includegraphics[width=3.4in,angle=0]{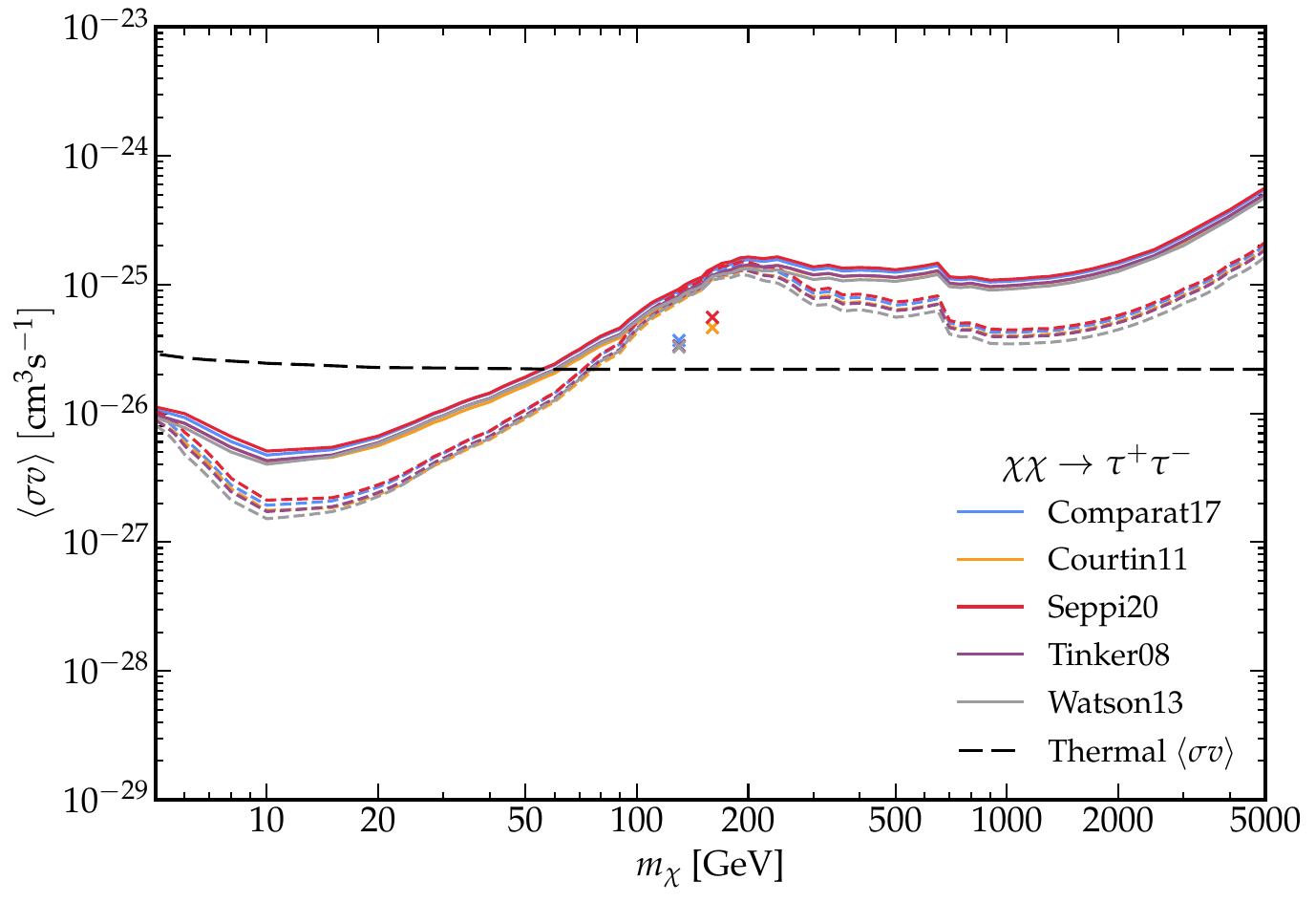}
\hspace{0.5cm}
\includegraphics[width=3.4in,angle=0]{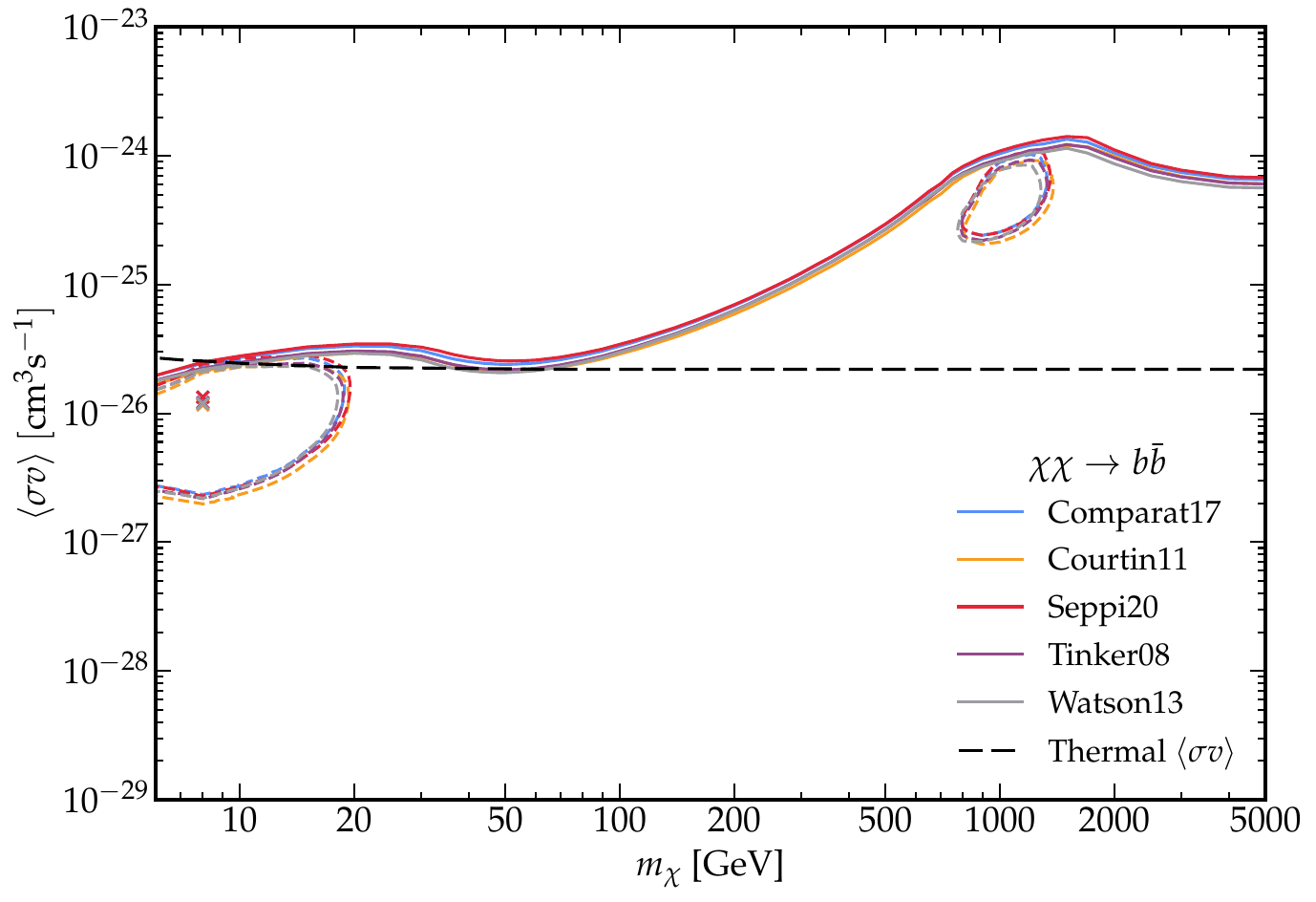}\\
\hspace{-0.2cm}
\includegraphics[width=3.4in,angle=0]{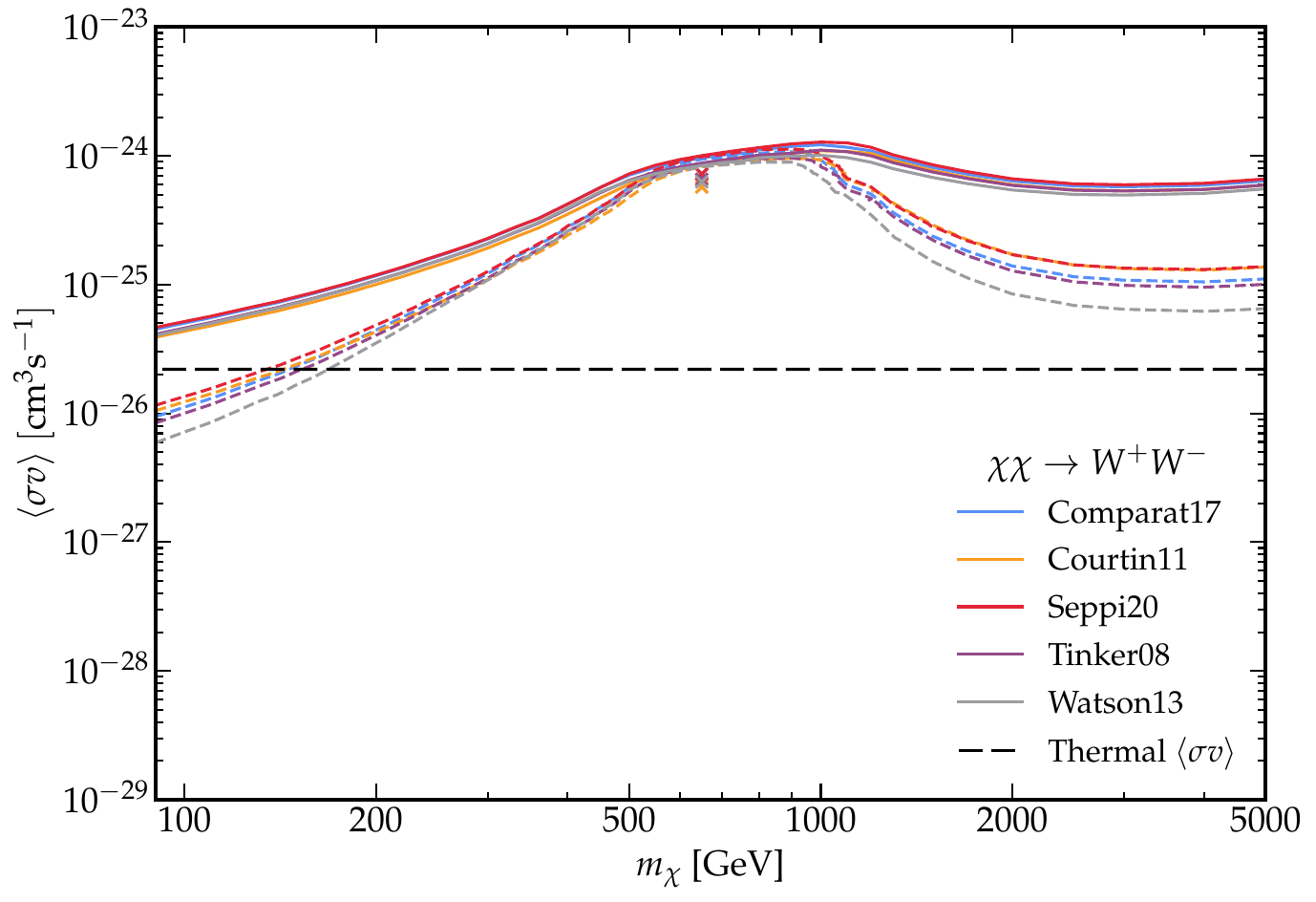}
\hspace{0.5cm}
\includegraphics[width=3.4in,angle=0]{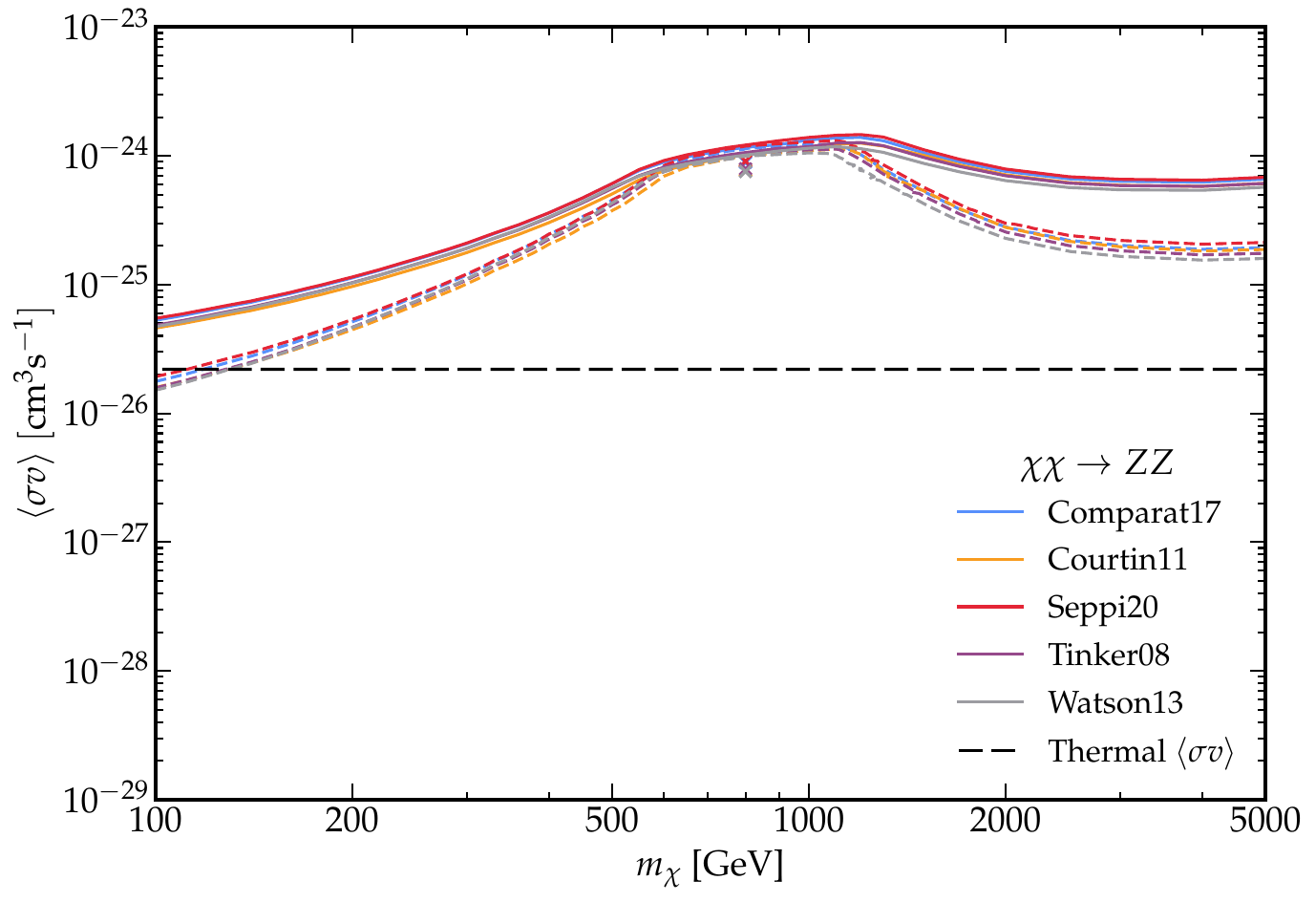}
\end{centering}
\vspace{-0.7cm}
\caption{
Using the IGRB model A spectrum, we show the impact of alternative halo mass functions on the derived dark matter properties. 
In Fig.~\ref{fig:DMlimits}, we used the Tinker08 HMF for the evaluation of the extragalactic dark matter flux. 
From that figure, we pick the best fit dark matter parameter point ``x'', the open or closed dashed line that gives the parameter space within $1-\sigma$ from that point and also the $95\%$ upper limit on the dark matter annihilation cross-section. 
The five colors represent the five choices of HMFs discussed in Section~\ref{subsec:Extragalactic_DM} and described in Refs.~\cite{2008ApJ...688..709T, 2011MNRAS.410.1911C, 2013MNRAS.433.1230W, 2017MNRAS.469.4157C, 2021A&A...652A.155S}.
The black dashed line gives the thermal relic annihilation cross-section. As with Fig.~\ref{fig:DMlimits}, from top to bottom and from left to right we show our results for the $\chi \chi \rightarrow \tau^{+} \tau^{-}$, $\chi \chi \rightarrow b\bar{b}$, 
$\chi \chi \rightarrow W^{+} W^{-}$ and $\chi \chi \rightarrow Z Z$ annihilation channels. 
Our results are affected very mildly from the use of alternative HMFs.}
\vspace{-0.6cm}
\label{fig:DMlimits_Alt_HMF}
\end{figure*}

One of the main uncertainties in evaluating the dark matter contribution to the IGRB, is the assumed halo mass function $dn/dM \,(M,z)$ of Eq.~\ref{eq:deltaSq}.
In Fig.~\ref{fig:DMlimits_Alt_HMF}, we show our results for five alternative HMFs for each of the four annihilation channels we study (see discussion in Section~\ref{subsec:Extragalactic_DM}).
Using alternative HMFs, has a very minimal effect on the location of the dark matter properties that give the overall best fit (alternative color ``x'' symbols), or on the dashed lines that give the parameter space within $1-\sigma$ from each of the respective color ``x'' points. Furthermore, the derived $95\%$ upper limits are marginally affected by the HMF assumption. 

The other major uncertainty in evaluating the dark matter contribution to the IGRB, is the boost factor $b_{\textrm{sh}}(M)$ to the annihilation signal, from substructures within a given halo (see Eq.~\ref{eq:deltaSq}).
However, alternative boost factor choices predominantly affect the overall normalization of the extragalactic dark matter flux component. 
As we show in Fig.~\ref{fig:DMcomponents}, the extragalactic component is the dominant dark matter term to the IGRB flux. 
Alternative boost factor assumptions would translate the quoted limits along the $\langle \sigma v \rangle$ axis without changing their shape 
\footnote{The extragalactic and galactic at high latitudes dark matter fluxes do not have identical spectral shapes as shown in Fig.~\ref{fig:DMcomponents}. However, under no reasonable assumptions on the boost factor, does the galactic term become dominant for dark matter with $m_{\chi} < 1$ TeV. For TeV scale dark matter particles where a big fraction of their extragalactic gamma-ray flux is attenuated, alternative boost factor assumptions can affect the shape of the $95\%$ upper limits. We consider that to be a minor effect that might be worth exploring further when the IGRB spectrum is much more accurately measured above 100 GeV.}. 
The model for the boost factor from Ref.~\cite{Gao:2011rf}, that we use here is by now the standard choice. 

\begin{figure*}
\begin{centering}
\hspace{-0.2cm}
\includegraphics[width=3.4in,angle=0]{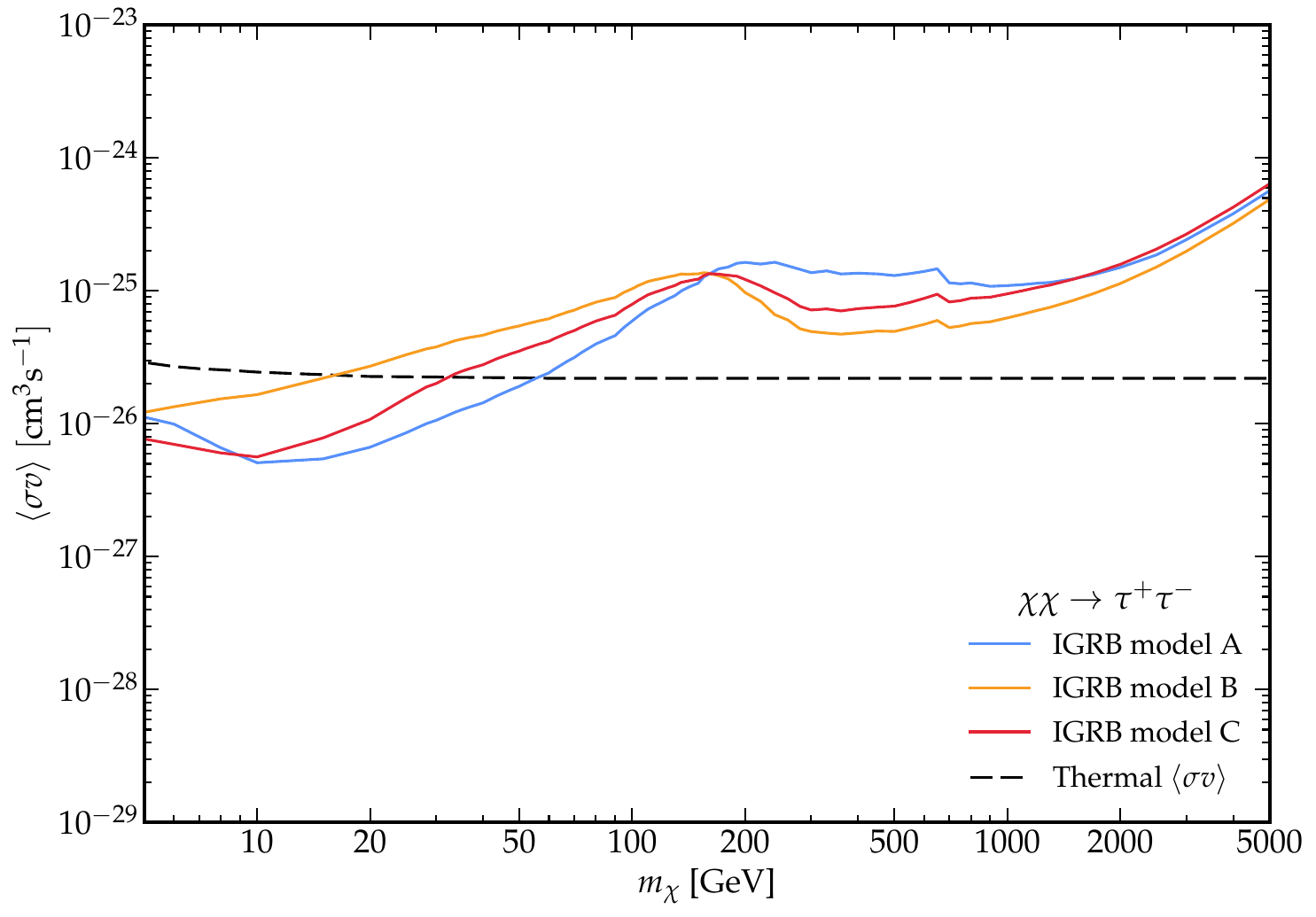}
\hspace{0.5cm}
\includegraphics[width=3.4in,angle=0]{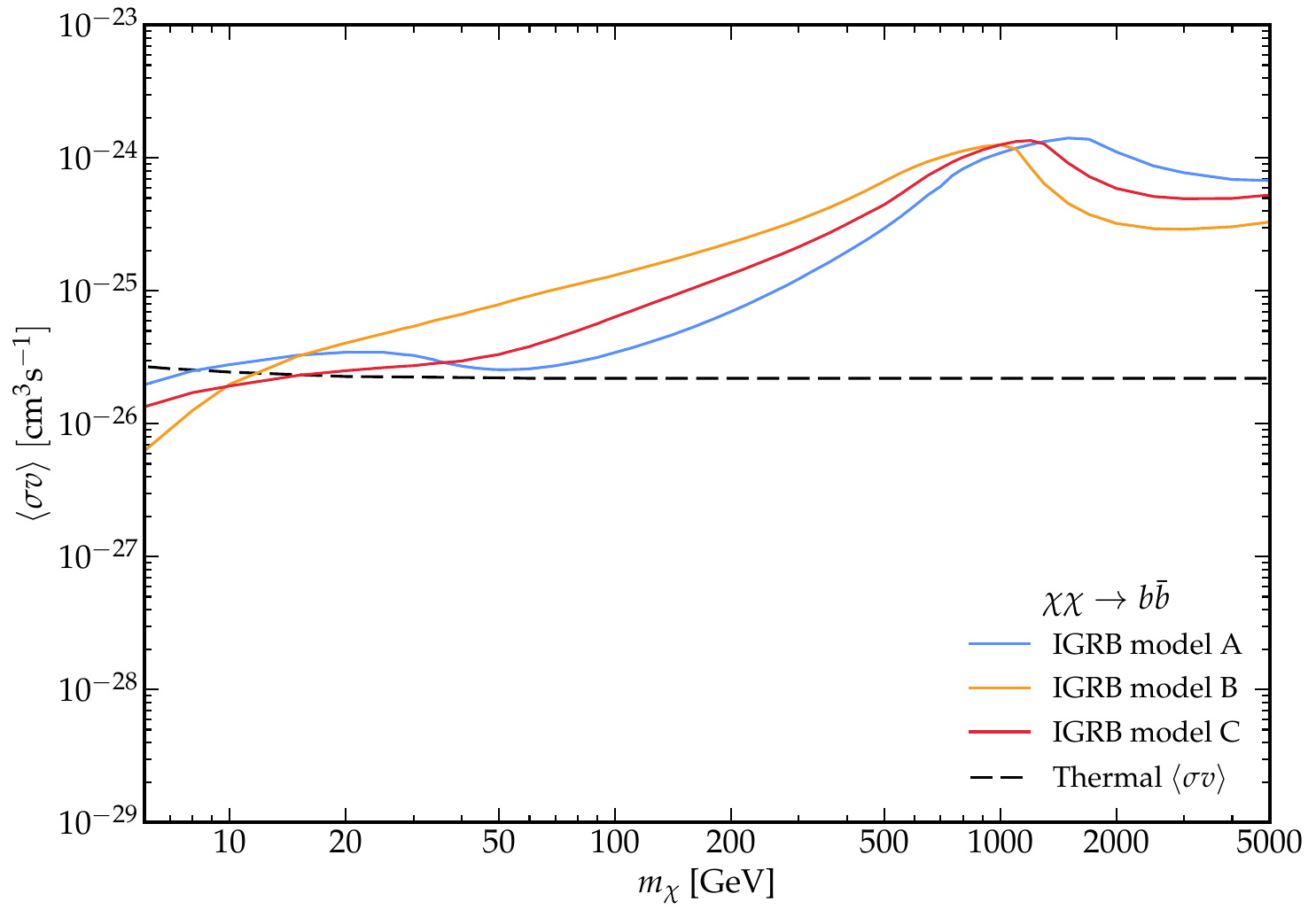}\\
\hspace{-0.2cm}
\includegraphics[width=3.4in,angle=0]{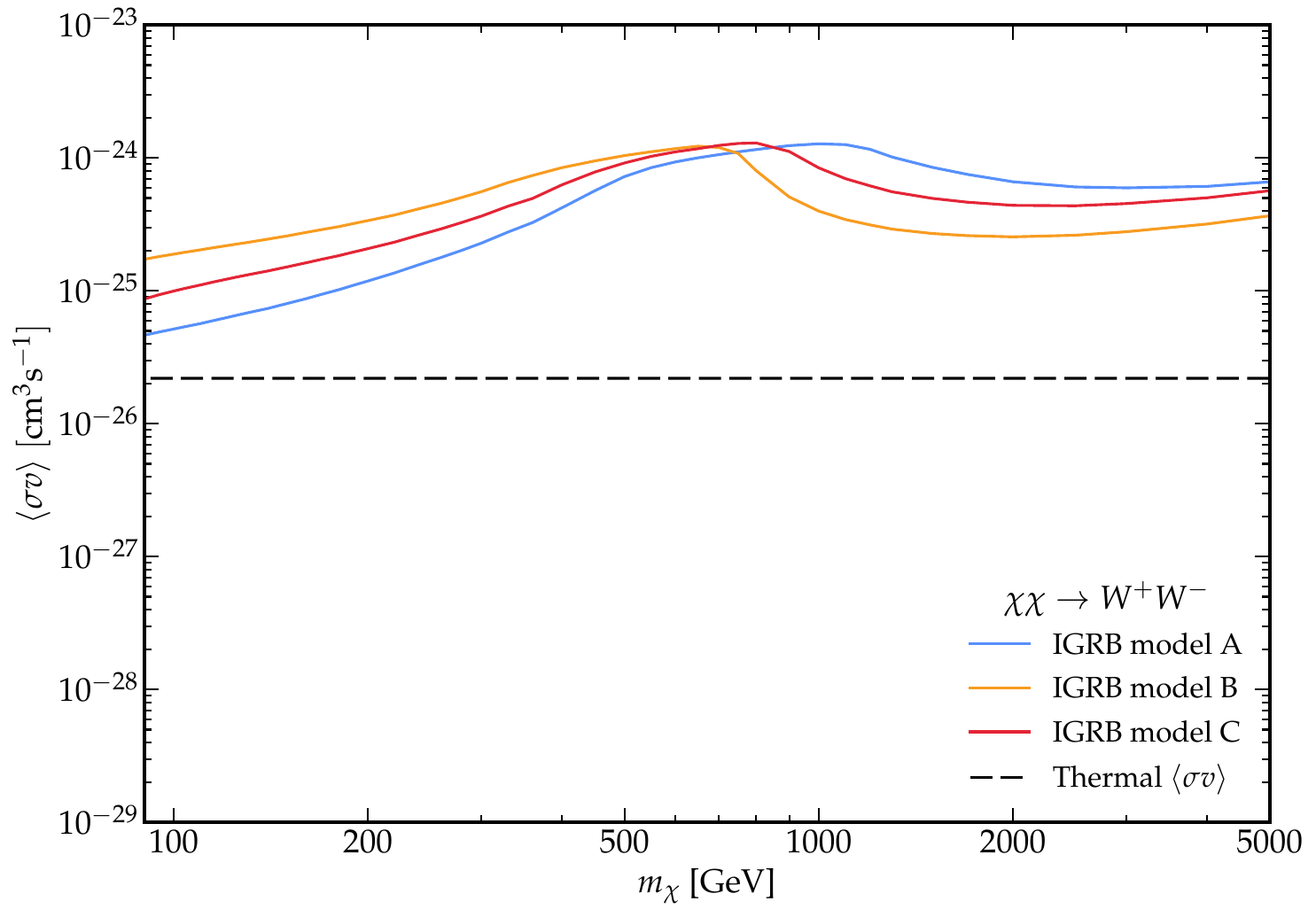}
\hspace{0.5cm}
\includegraphics[width=3.4in,angle=0]{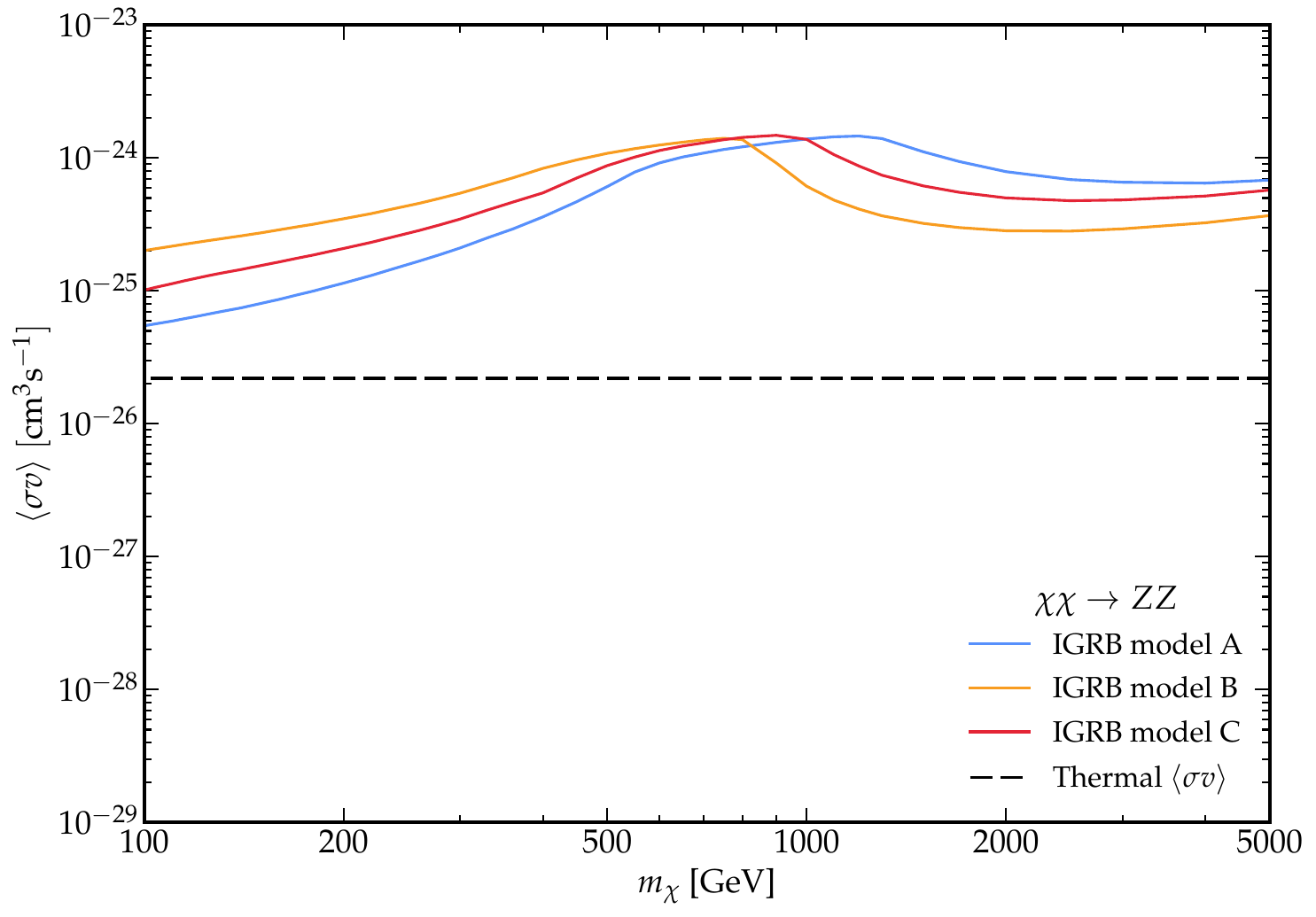}
\end{centering}
\vspace{-0.7cm}
\caption{The $95\%$ upper limits on the dark matter annihilation cross-section, as derived using the IGRB model A spectrum (blue line), -also shown in Fig.~\ref{fig:DMlimits}-, the IGRB model B spectrum (orange line) and the IGRB model C spectrum (red line). The four panels give results for the same four annihilation channels as Figs.~\ref{fig:DMlimits} and~\ref{fig:DMlimits_Alt_HMF}. We note that within a factor of $\simeq 2$ the dark matter limits from the three IGRB spectral models agree.}
\vspace{-0.6cm}
\label{fig:DMlimits_Alt_IGRB}
\end{figure*}

In Fig.~\ref{fig:DMlimits_Alt_IGRB}, we explore how our $95 \%$ upper limits change by using the alternative IGRB spectrum evaluations from the \textit{Fermi}-LAT collaboration in Ref.~\cite{Fermi-LAT:2014ryh}. 
These alternative IGRB spectra exist, as even at the highest galactic latitudes, the total gamma-ray flux is mostly due to either the galactic emission or identified point sources. 
Subtracting those components to get the IGRB comes with its modeling/systematic uncertainties. 
We find that within a factor of $\simeq 2$, our dark matter limits are in agreement. The main difference between the limit lines is that they appear to be
somewhat transposed with respect to each other.

The IGRB spectra of Ref.~\cite{Fermi-LAT:2014ryh}, were derived using the first 50 months of \textit{Fermi}-LAT observations, when now we are at the 17th year of data taking. 
A better measurement of the IGRB spectrum is feasible, both by having statistically smaller errors and also possibly reducing the systematic errors associated to modeling (and subsequently subtracting) the contribution of the Milky Way's diffuse emission at high latitudes.  
A more accurate IGRB spectrum, will allow us to derive more stringent limits on the properties of dark matter and also address if there is a need for a dark matter contribution to the data.
 
\section{Discussion and Conclusions}
\label{sec:conclusions}

In this paper we revisit the limits that the IGRB measurement can place on dark matter annihilation. 
Relying on recent developments on the modeling of the spectra, luminosity distribution and redshift distribution properties of the conventional background astrophysical sources of the IGRB, we build models for the contribution of each component to the IGRB. 
In particular, in our analysis we separately model the contribution to the IGRB from BL Lacs and FSRQs, discussed in Section~\ref{subsec:BLLac-FSRQ}. We also account for the gamma-ray emission from unresolved star-forming and starburst galaxies, as described in Section~\ref{subsec:Starforming}.
Radio galaxies, discussed in Section~\ref{subsec:Radio}, is another potentially significant component to the IGRB. 
At the high end of the IGRB spectrum, we expect and in fact confirm, that ultra-high-energy cosmic rays interacting with the intergalactic infrared background are a major component of the IGRB (see discussion of Section~\ref{subsec:UHECRs}). 
Finally, while a minor component to the IGRB, we include the contribution of high-latitude unresolved millisecond pulsars, as modeled in Section~\ref{subsec:MSPs}.   

We then combine these known astrophysical components to explain the IGRB spectrum. 
The IGRB spectrum provided by the \textit{Fermi} collaboration~\cite{Fermi-LAT:2014ryh}, comes with its own modeling uncertainties, mostly related to properly accounting for and subtracting the diffuse emission from the Milky Way, in the region of the sky used to measure the IGRB gamma-ray flux. Ref.~\cite{Fermi-LAT:2014ryh}, has provided three spectral models for the IGRB, referred to as models A, B and C. We use all three spectra.

In our analysis we fit the normalizations of the known astrophysical components and also allow for some freedom in the combined spectral index of each population of unresolved extragalactic sources, i.e. the BL Lacs, FSRGs, star-forming $\&$ starburst galaxies and the radio galaxies. 
That freedom is used to both account for the modeling uncertainties on the properties of each of those populations and also is important in studying the IGRB as probe for dark matter.
We discuss the details of the fitting procedure in Sections~\ref{subsec:Combination} and~\ref{subsec:analysis} and show our best fit result in Fig.~\ref{fig:BackModA} for the IGRB spectrum model A. 
Assuming a very narrow range in the freedom of the spectra of the extragalactic sources does not give a very good good fit. However, once allowing for some greater freedom in the combined spectral indices of these sources, we find good fits of $\chi^{2}$/dof $\simeq 1$ to the IGRB spectral data. 

Subsequently, we include the possible contribution from dark matter annihilation to the IGRB. In modeling the dark matter component, we account for both the dominant extragalactic emission, i.e. the emission of gamma rays from distant galaxies/halos (and their substructure), and  also account for the gamma-rays produced from dark matter annihilations occurring at high latitudes but within the Milky Way's dark matter halo. For the extragalactic dark matter component we also study alternative models on the halo mass function that describes the mass and redshift distribution of dark matter halos. Our modeling of the dark matter component to the IGRB is detailed in Section~\ref{sec:DM}.

We find that including a dark matter component to the IGRB, can improve the fit to the spectral data. 
The importance of such a component depends on the dark matter annihilation channel, but is less than $2 \sigma$, mostly seen for the $\chi \chi \rightarrow b \bar{b}$ channel. 
We note however that in checking for a possible dark matter component, we have used a great amount of freedom in marginalizing over the normalization and spectral indices of the conventional astrophysical components. 
With an analysis of more lengthy observations of the gamma-ray sky than those used in Ref.~\cite{Fermi-LAT:2014ryh}, we will be able to check if this possible hint of a dark matter component will become more statistically significant (as it should if true) or is some spectral artefact related to how the IGRB is evaluated. 
At the time this paper is written there are about four times more lengthy observations of the gamma-ray sky. A reanalysis of the IGRB spectrum will be of great benefit. 

We derive upper limits on the annihilation cross-section of dark matter particles with masses from 5 GeV to 5 TeV annihilating into four district channels: i) $\chi \chi \rightarrow \tau^{+} \tau^{-}$ ii) $\chi \chi \rightarrow b\bar{b}$, iii) $\chi \chi \rightarrow W^{+} W^{-}$ and iv) $\chi \chi \rightarrow Z Z$. Our limits are given in Figs.~\ref{fig:DMlimits},~\ref{fig:DMlimits_Alt_HMF},~\ref{fig:DMlimits_Alt_IGRB},~\ref{fig:DMlimits_IGRB_B} and~\ref{fig:DMlimits_IGRB_C}. We find that the limits from the IGRB spectra are very robust to alternative halo mass function assumptions (at the $O(0.1)$ level). 
More importantly, the limits on the annihilation cross-section depend within a factor of $\simeq 2$ on the IGRB spectral model used (model A vs B vs C).  
A better understanding of the galactic diffuse emission at high latitudes would allow us to reduce the underlying systematic uncertainties of the IGRB spectrum. This is another main improvement on making the IGRB measurement a stronger probe for dark matter. 

\begin{figure}
\begin{centering}
\hspace{-0.3cm}
\includegraphics[width=3.35in,angle=0]{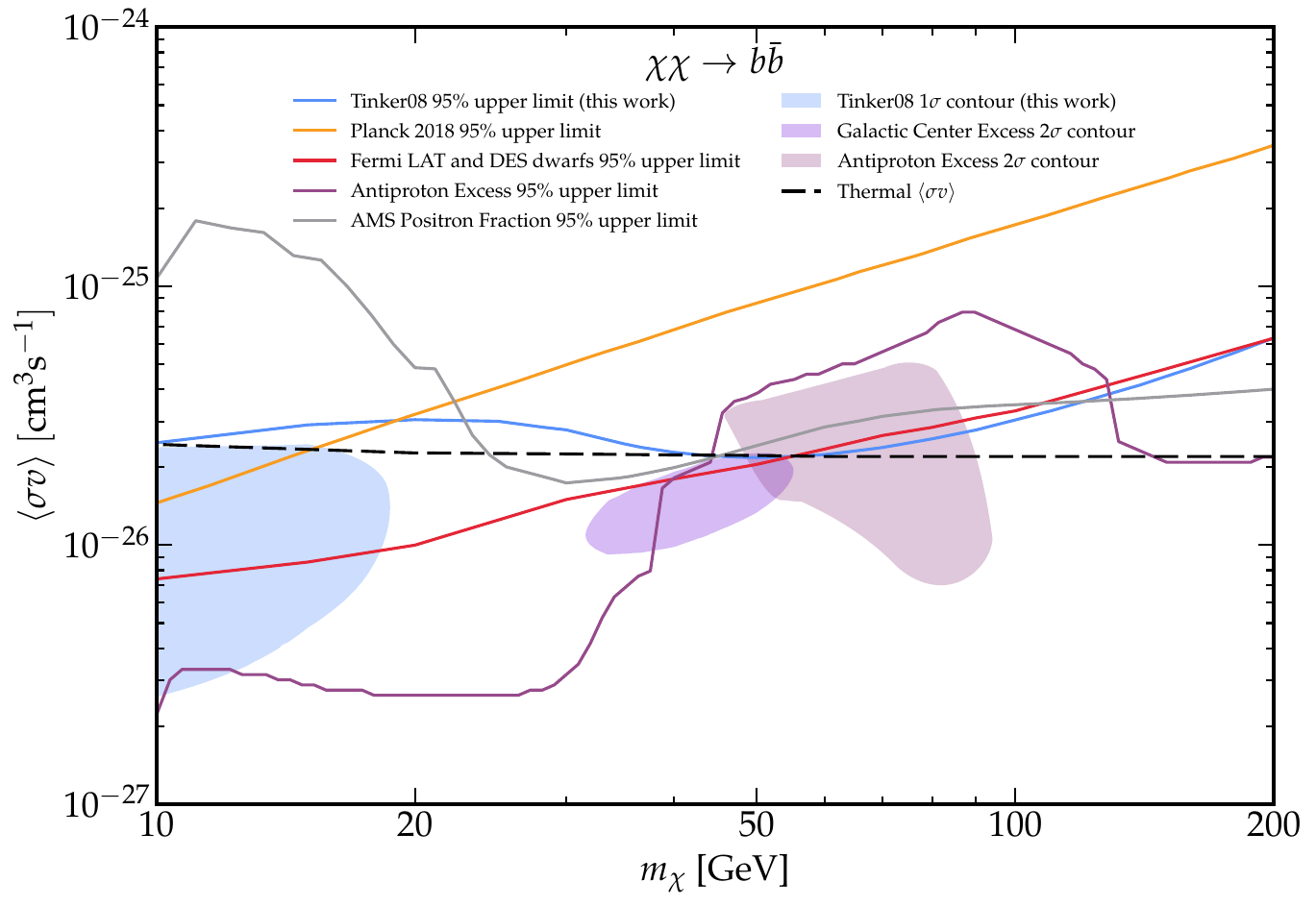}
\end{centering}
\vspace{-0.4cm}
\caption{
Comparison of this work's limits on dark matter (blue line) and hint for an excess (blue region) and other limits from the CMB \cite{Planck:2018vyg}, the \textit{Fermi} dwarf galaxies \cite{Fermi-LAT:2016uux}, the cosmic-ray antiprotons (limits and excess from Ref.\cite{Cholis:2019ejx}) and from cosmic-ray positrons \cite{Krommydas:2022loe}. The GCE (from Ref.~\cite{Cholis:2021rpp}) and the antiproton excess are also shown with their respective $2\sigma$ contours.}
\vspace{-0.7cm}
\label{fig:Limits_and_Excesses}
\end{figure}

Even with the above mentioned caveats, the IGRB limits on dark matter are similar in strength to those derived by dwarf spheroidal galaxy observations and cosmic-ray antiprotons. Moreover, the dark matter parameter space needed to explain the gamma-ray galactic center excess and the cosmic-ray antiproton excess can not be excluded by our limits. 
In Fig.~\ref{fig:Limits_and_Excesses}, for the case where dark matter annihilates to $b\bar{b}$ quarks, we show how the limits from this work compare to the gamma-ray galactic center excess, the cosmic-ray antiproton excess, and limits from dwarf spheroidal galaxies, the cosmic microwave background and cosmic-ray antiprotons and positrons\footnote{The limit from cosmic-ray positrons depends on uncertainties in the properties of the Milky Way pulsars population, the interstellar medium properties, assumptions on cosmic-ray secondary and primary species and on solar modulation uncertainties. We present here  the lower(tighter) end of 95$\%$ CL limit (see Ref.~\cite{Krommydas:2022loe} for details).}. 

As a final note, we have tested the sensitivity of our dark matter limits on the freedom we use in marginalizing over the normalizations and spectra of the conventional astrophysical components and find that our limits on the annihilation cross-section are robust at the $O(0.01)$ level.

As part of this work we make publicly available our dark matter and astrophysical background simulation spectral files in \texttt{https://zenodo.org/records/13351935}. 
  
\acknowledgements
We thank Dan Hooper and Tim Linden for useful discussions.
We acknowledge the use of \texttt{GALPROP} \cite{GALPROPSite, galprop}
and the \path{Python} \cite{10.5555/1593511} modules, \texttt{COLOSSUS} \cite{2018ApJS..239...35D, COLOSSUSwebsite}, \path{numpy} \cite{harris2020array},
\path{SciPy} \cite{2020SciPy-NMeth}, \path{matplotlib} \cite{Hunter:2007}, \path{Jupyter} \cite{Kluyver2016jupyter}, and \path{iminuit} \cite{iminuit,James:1975dr}.
IC acknowledges that this material is based upon work supported by the U.S. Department of Energy, Office of Science, 
Office of High Energy Physics, under Award No. DE-SC0022352.

\begin{appendix}

\section{The impact of background modeling freedom on the dark matter limits}
\label{app:background_scrict_fit}

In probing for a possible dark matter contribution to the IGRB spectrum, we have accounted for and marginalized over the conventional astrophysical background components.
Those components are the combined emission from unresolved BL Lacs, FSRQs, star-forming and starburst galaxies and radio galaxies. Also, we account for the interaction of ultra-high-energy cosmic-ray nuclei scattering with the infrared background and for unresolved galactic MSPs at high latitudes.   
Our base assumptions in modeling those astrophysical sources of IGRB emission are described in Section~\ref{sec:astro_backgrounds}. 
In fitting the IGRB spectrum, with and without dark matter, 
we allow -from the base assumptions-, for an additional freedom in the normalization of each one of these six components and also for a pivot of the spectral index $\Delta \gamma$, of the four unresolved populations of galaxies. 
The exact range is given in Table~\ref{tab:FitFreedom}. 

We have noticed that such a freedom allows for good fits to the data with $\chi^{2}$/dof $\lsim 1$. 
However, it also leads to a great amount of degeneracy between these components. In some cases this freedom allows for two astrophysical components to have after fitting, nearly identical spectral indices and thus highly anti-correlated normalizations. 
Such an example can be seen in Fig.~\ref{fig:BestFitExamples}, (bottom two panels), where after the fit is performed the flux component from start-forming galaxies and from FSRQs have essentially the same spectral shape above 0.5 GeV. 
In  such a case increasing the normalization of one component reduces the normalization of the other without any change in the quality of the fit. 

We repeat that in probing for an additional dark matter contribution to the IGRB, a large freedom in modeling the background contribution is the conservative approach. 
We test here reduced ranges of fit normalizations and spectral shape changes $\Delta \gamma$. The more strict range of fitting parameters is given in Table~\ref{tab:FitFreedom_strict}.
Such a choice, allows for breaking the degeneracy between the background astrophysical components \footnote{With the restricted parameter range we used for the BL Lac population a value of $\mu^{*} = 2.1$.}.

\begin{table}[ht]
    \begin{tabular}{ccc}
         \hline
           Component &  Normalization Range & $\Delta{\gamma}$ \\
            \hline \hline
            BLLac &  [0.3, 1.0] & [-0.1, + 0.1] \\   
            SFRQ &  [0.5 1.0] & [-0.04, + 0.03] \\
            SF \& SB & [0.1, 1.0] & [-0.2, +0.2] \\
            RG &  [0.2, 1.0] & [-0.2, +0.2] \\
            UHECR &  [0.2, 5] & 0.0 \\     
            MSP &  [0.5, 1.5] & 0.0 \\
        \hline \hline 
        \end{tabular}
        \vspace{-0.3cm}
\caption{As in Table~\ref{tab:FitFreedom}, we give the reduced ranges of freedom in the normalization and spectral shape of each background astrophysical component when fitting to the IGRB spectrum.} 
\vspace{-0.5cm}
\label{tab:FitFreedom_strict}
\end{table}

\begin{figure}
\begin{centering}
\hspace{-0.0cm}
\includegraphics[width=3.6in,angle=0]{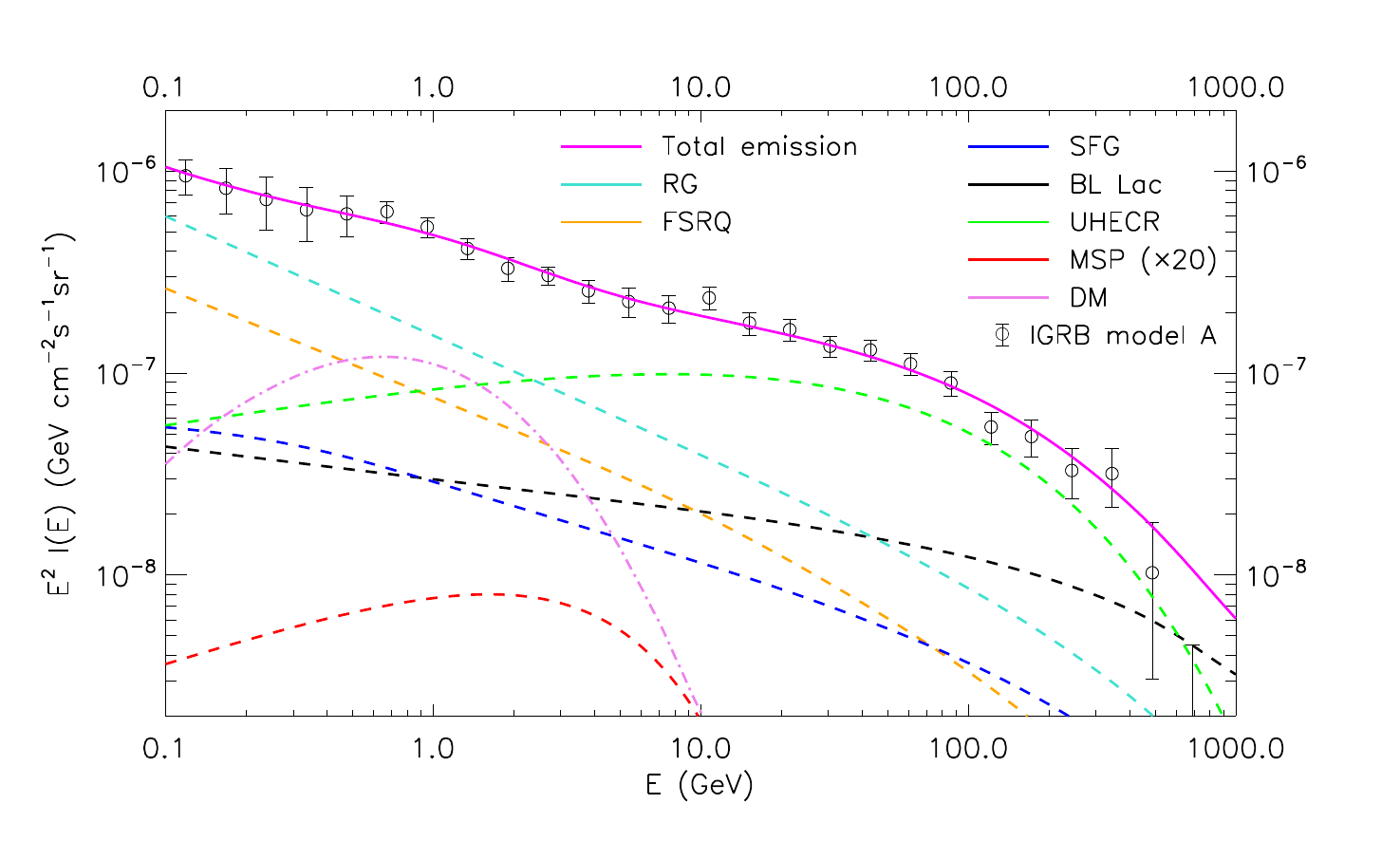}\\
\includegraphics[width=3.6in,angle=0]{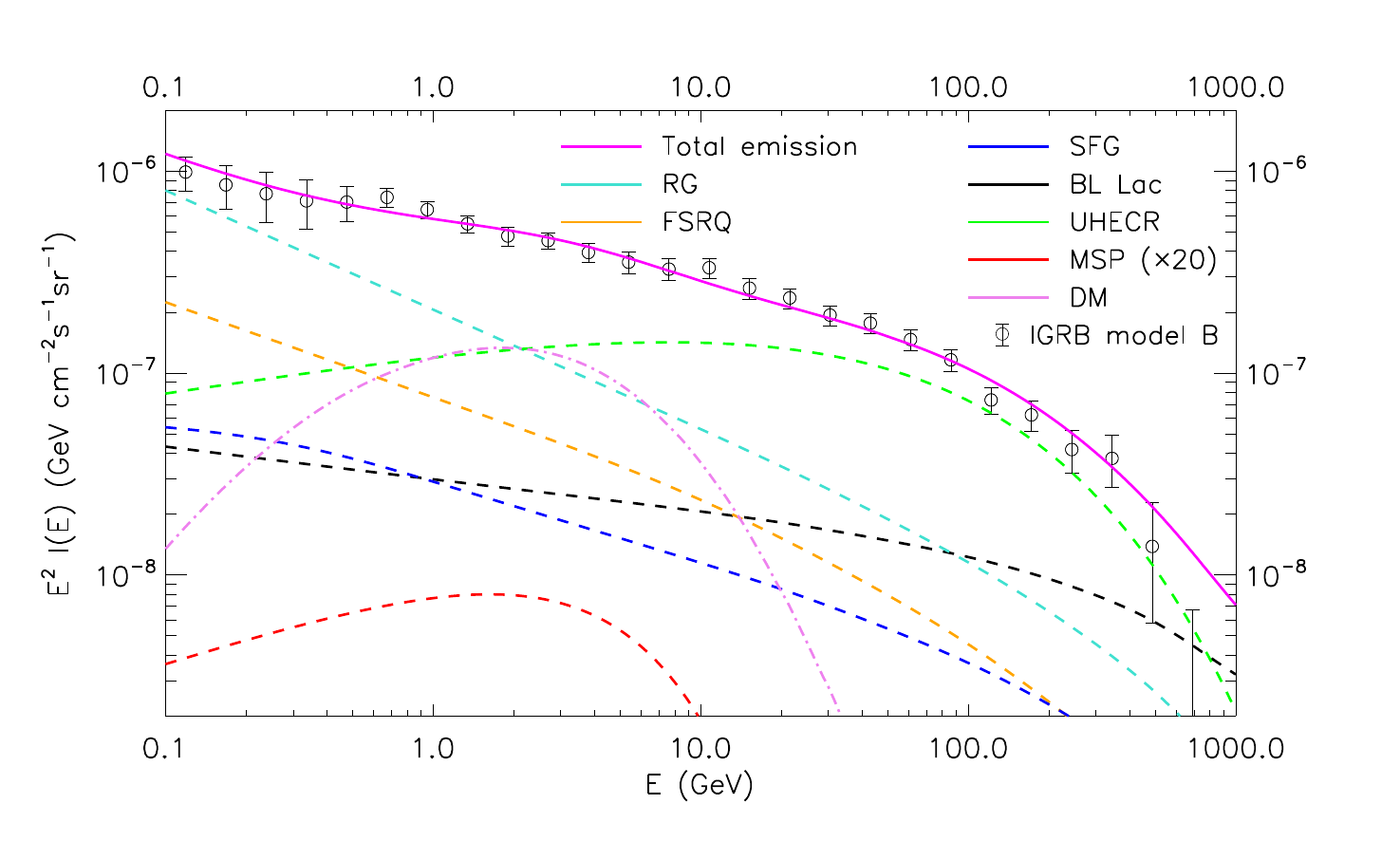}\\
\includegraphics[width=3.6in,angle=0]{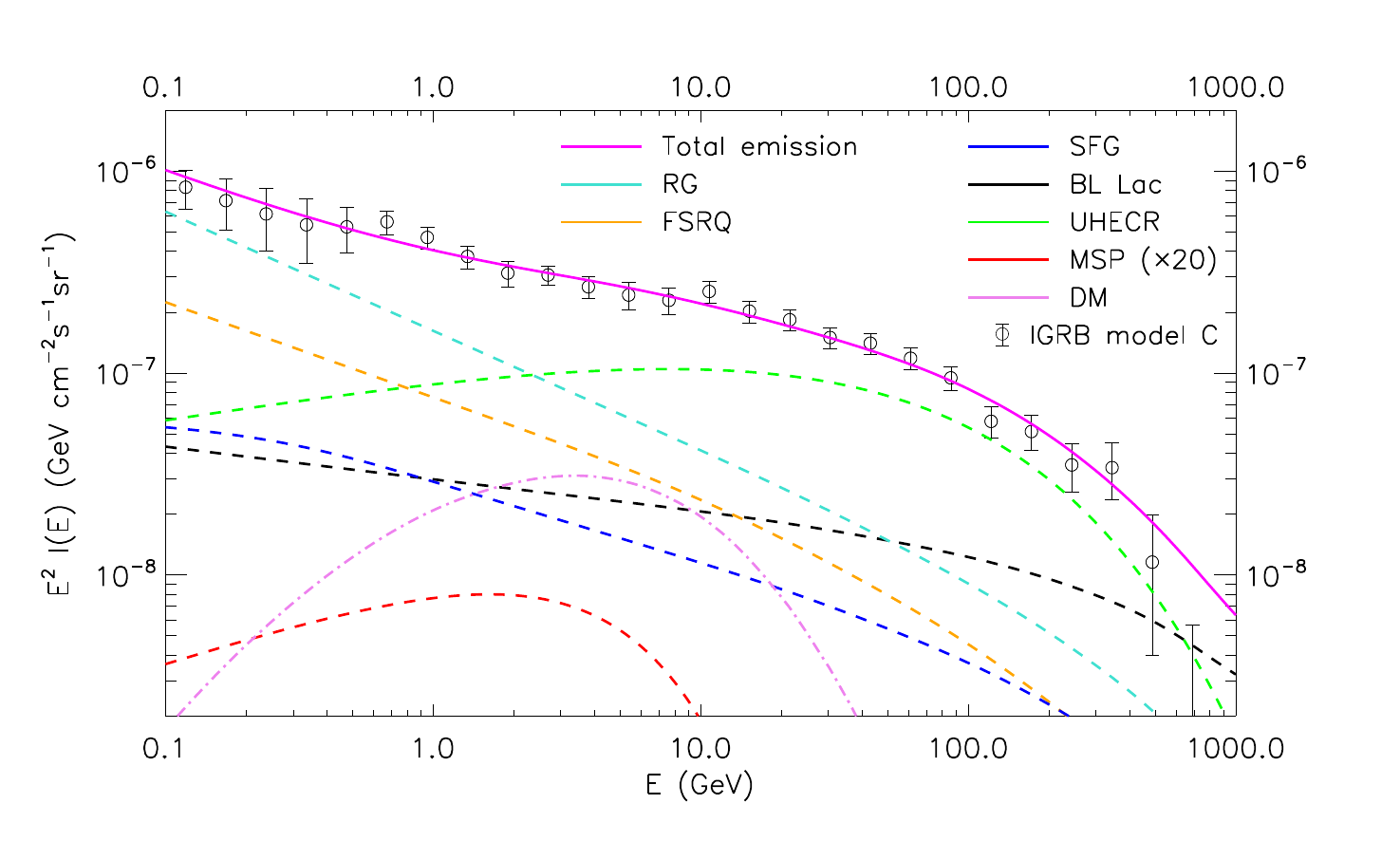}
\end{centering}
\vspace{-0.7cm}
\caption{
Like with Fig.~\ref{fig:BestFitExamples}, the contribution of all components to the IGRB. Here we used the more constrained fit ranges for the background components (see text for details).
The top panel shows the IGRB spectrum A for a 15 GeV dark matter particle annihilating to $b\bar{b}$ quarks, with a cross-section of $\sigma v = 1.5\times 10^{-26} \textrm{cm}^3/\textrm{s}$, giving a $\chi^{2}$/dof = 1.05. In the middle panel we show the IGRB spectrum B, for a 50 GeV particle annihilating to $b\bar{b}$, with $\sigma v = 5.9\times 10^{-26} \textrm{cm}^3/\textrm{s}$; leading to a $\chi^{2}$/dof = 1.26.
In the bottom panel we show IGRB model C, for a 100 GeV  particle, with $\sigma v = 2.9\times 10^{-26} \textrm{cm}^3/\textrm{s}$; giving a $\chi^{2}$/dof = 1.06.}
\vspace{-0.6cm}
\label{fig:BestFitExamples_strict}
\end{figure}

In Fig.~\ref{fig:BestFitExamples_strict}, we show our results with this fitting procedure. For simplicity and ease in comparison, we use the same combination of IGRB spectra (A, B and C) and specific dark matter masses for the $\chi \chi \rightarrow b \bar{b}$ channel as shown in Fig.~\ref{fig:BestFitExamples}. 
In these panels the background components have more separable spectral shapes and somewhat different normalizations than we got when using the greater fitting freedom in the main body of this work.
This more strict fitting also leads to worse fits in $\chi^{2}$/dof. For these three examples from 0.7-0.8 that we had with our standard fitting assumptions to 1.0-1.3. 

Yet, we find that the reduced freedom in the parameter scan has a negligible effect in the derived $95\%$ upper limits on the dark matter annihilation cross-section versus mass for all channels. In fact, our results in Figs.~\ref{fig:DMlimits},~\ref{fig:DMlimits_Alt_HMF} and~\ref{fig:DMlimits_Alt_IGRB} are practically the same. The difference in the limits is at the $O(10^{-2})$ level. 

\section{Dark matter limits using the IGRB spectra B and C}
\label{app:IGRB_B_and_C}

In Fig.~\ref{fig:DMlimits} of the main body, we showed our dark matter fit results of the IGRB model A spectrum. 
Since models B and C are not considered inferior estimates of the IGRB spectrum in the analysis of Ref.~\cite{Fermi-LAT:2014ryh}, in this appendix we show the equivalent results to Fig.~\ref{fig:DMlimits} for these alternative spectra.

\begin{figure*}
\begin{centering}
\hspace{-0.2cm}
\includegraphics[width=3.4in,angle=0]{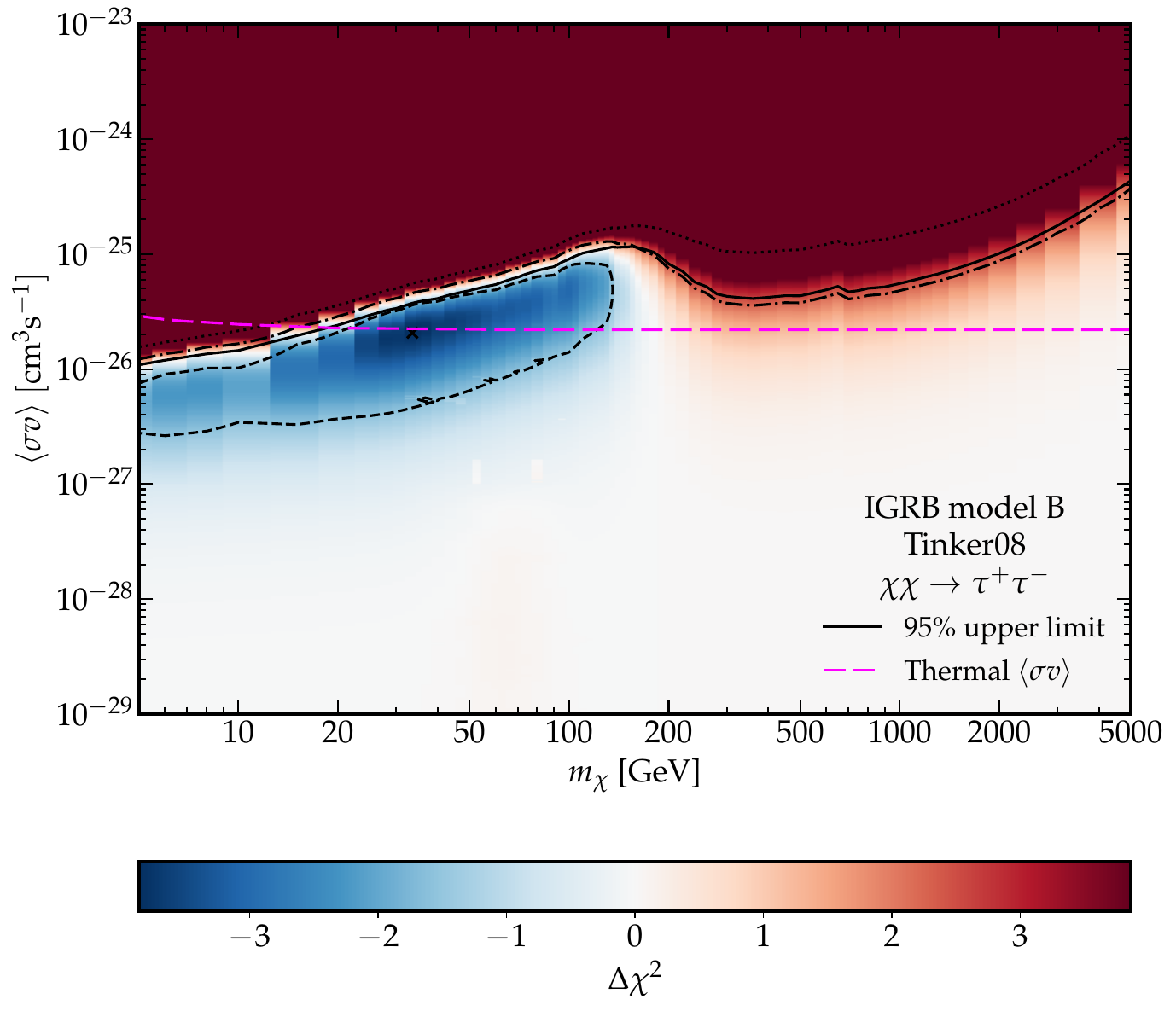}
\hspace{0.5cm}
\includegraphics[width=3.4in,angle=0]{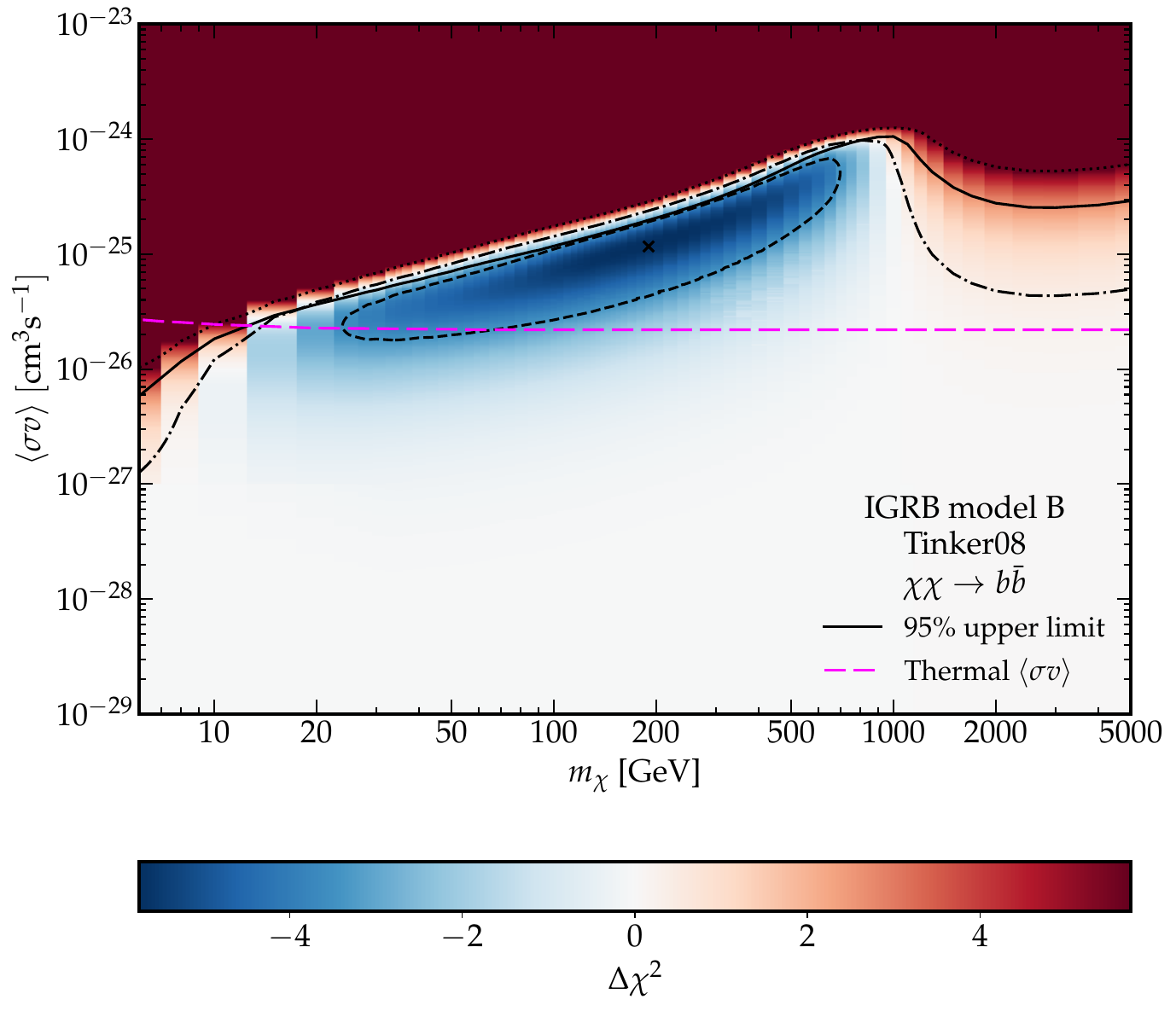}\\
\hspace{-0.2cm}
\includegraphics[width=3.4in,angle=0]{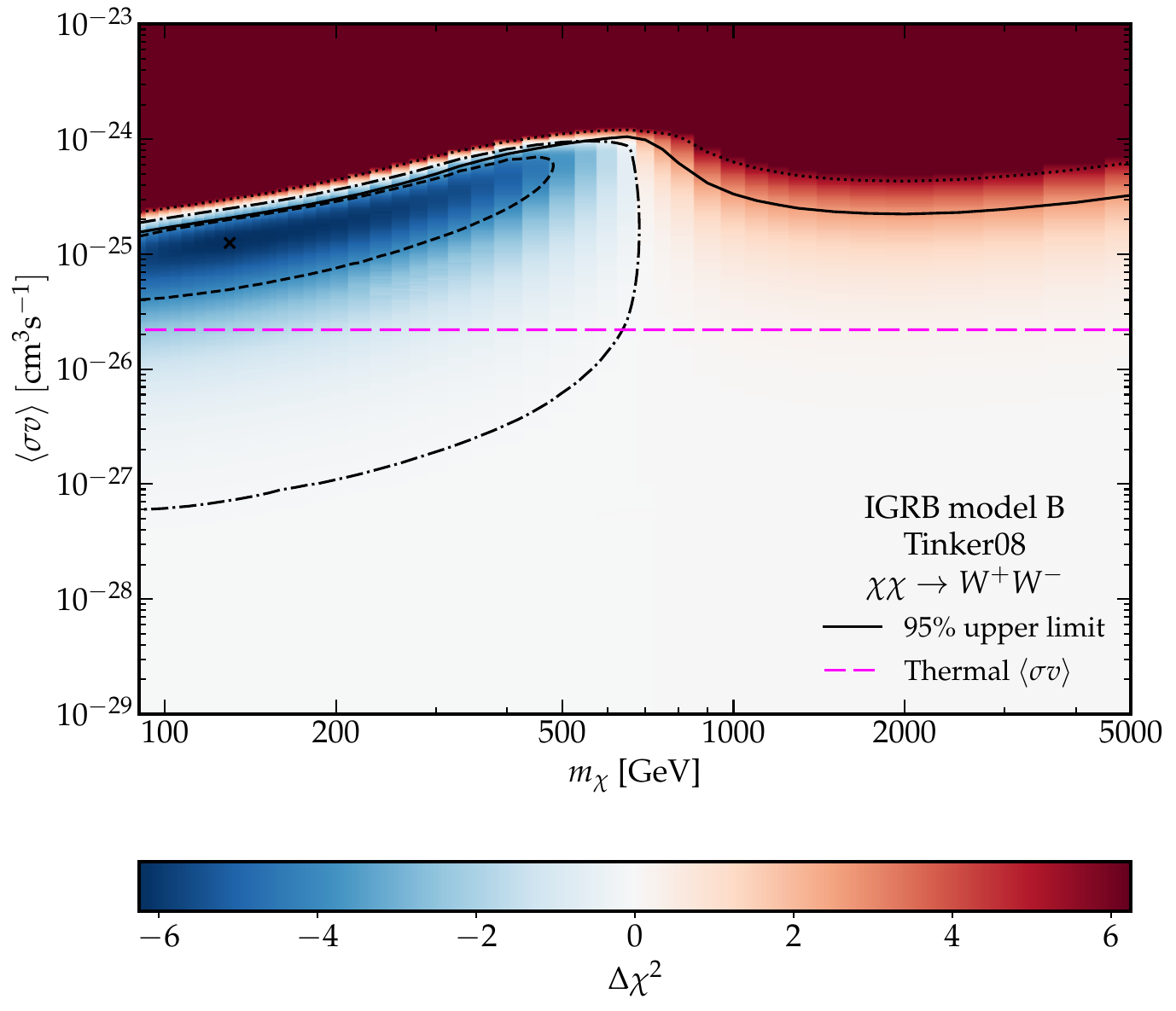}
\hspace{0.5cm}
\includegraphics[width=3.4in,angle=0]{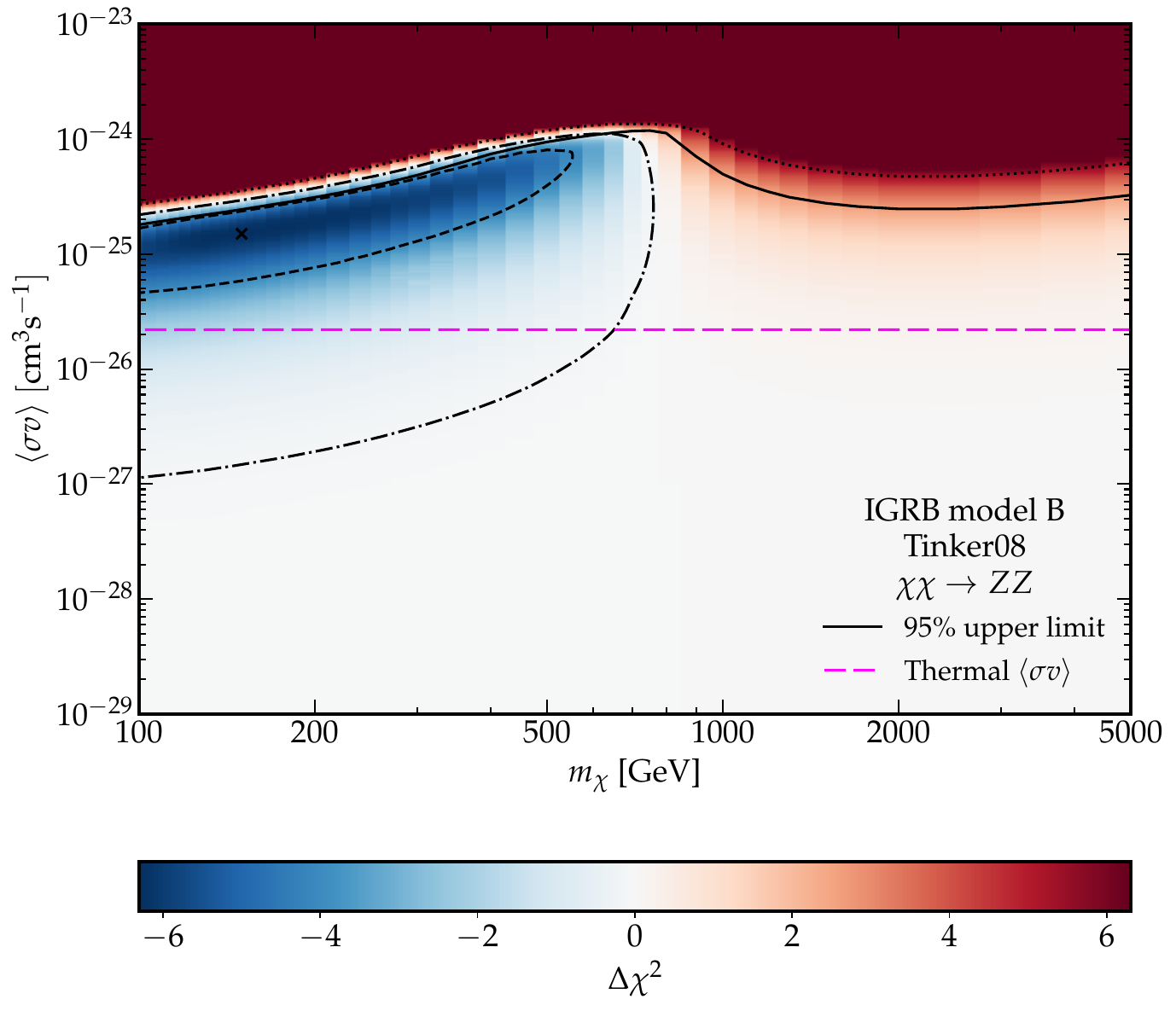}
\end{centering}
\vspace{-0.7cm}
\caption{As in Fig.~\ref{fig:DMlimits},
using instead IGRB model B spectrum. }
\vspace{-0.6cm}
\label{fig:DMlimits_IGRB_B}
\end{figure*}

\begin{figure*}
\begin{centering}
\hspace{-0.2cm}
\includegraphics[width=3.4in,angle=0]{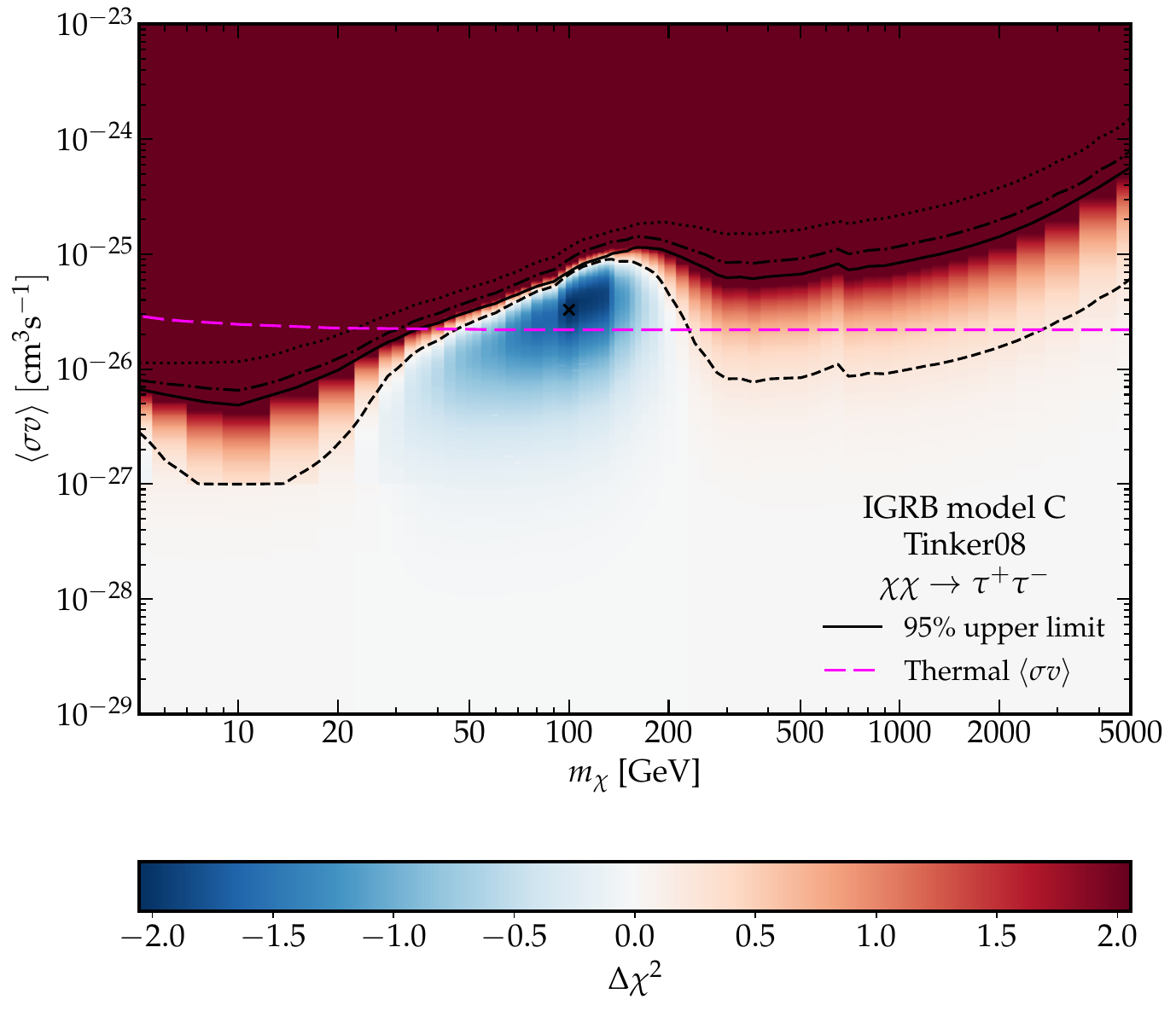}
\hspace{0.5cm}
\includegraphics[width=3.4in,angle=0]{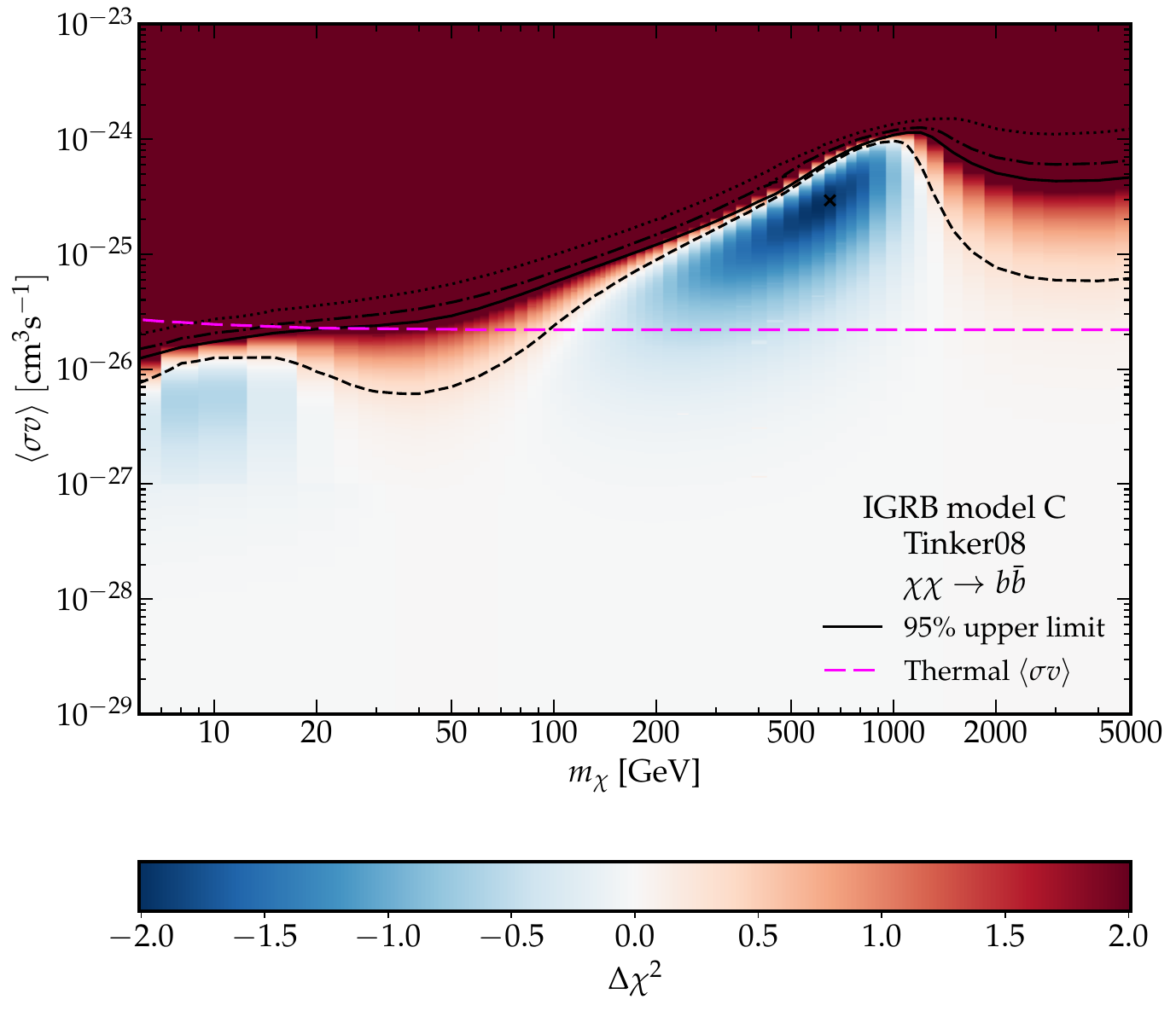}\\
\hspace{-0.2cm}
\includegraphics[width=3.4in,angle=0]{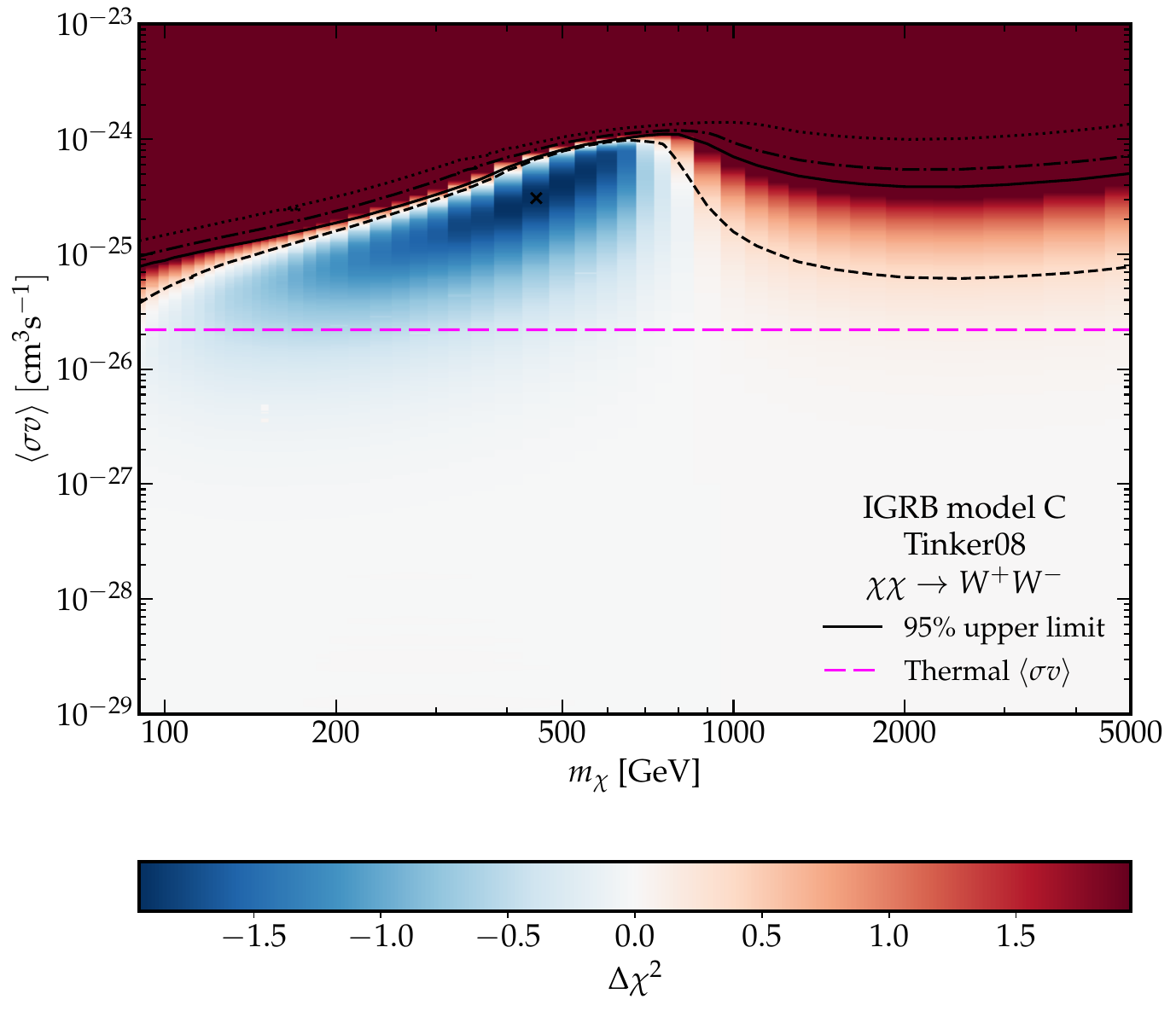}
\hspace{0.5cm}
\includegraphics[width=3.4in,angle=0]{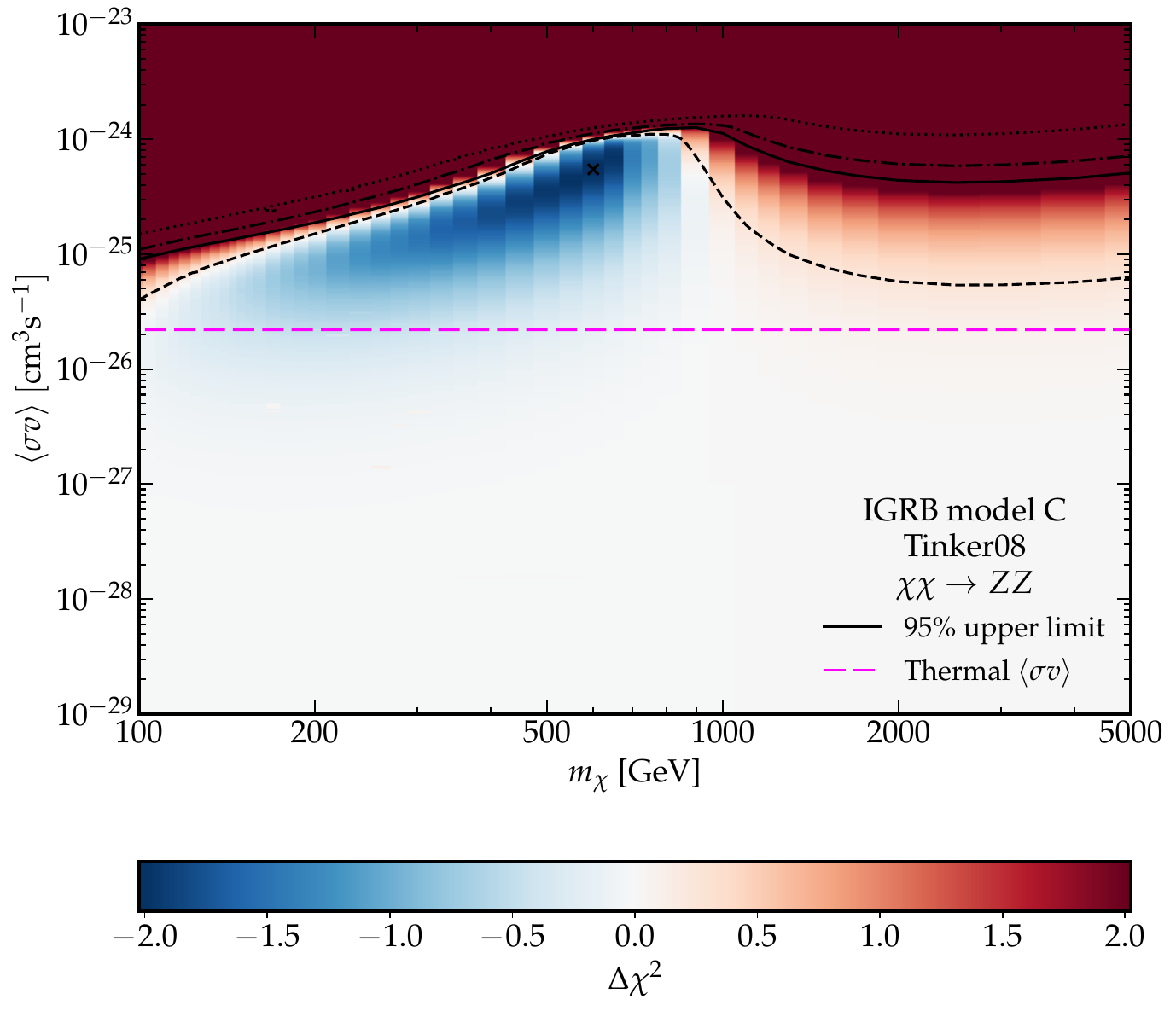}
\end{centering}
\vspace{-0.7cm}
\caption{As in Figs.~\ref{fig:DMlimits} and~\ref{fig:DMlimits_IGRB_B},
using instead IGRB model C spectrum.}
\vspace{-0.6cm}
\label{fig:DMlimits_IGRB_C}
\end{figure*}

Fig.~\ref{fig:DMlimits_IGRB_B}, shows our results using model B and Fig.~\ref{fig:DMlimits_IGRB_C}, model C. We show the exact same mass ranges and annihilation channels as in Fig.~\ref{fig:DMlimits}. 

The $95\%$ upper limits to the annihilation cross-section, for any given channel are also given in 
Fig.~\ref{fig:DMlimits_Alt_IGRB} of the main body and discussed there.

\end{appendix}  

\bibliography{EGRB_DM}

\end{document}